\documentclass[aps,prd,twocolumn,groupedaddress, showkeys, showpacs]{revtex4}
\pdfoutput=1
\usepackage{graphicx,subfigure,enumerate,pstricks,latexsym,amssymb}
\usepackage[english]{babel}

\providecommand{\dif}{\mathrm{d}} 
\def\beq{\begin{equation}}
\def\eeq{\end{equation}}
\def\bea{\begin{eqnarray}}
\def\eea{\end{eqnarray}}

\def\nn{\nonumber}

\newcommand{\nnd}{\discretionary{--}{--}{--}}
\newcommand{\Schw}{Schwarzschild}
\newcommand{\dS}{de~Sitter}

\def\d{\dif}
\def\der{|}

\def\SS{\Sigma}

\def\tens{\mu}
\def\xx{x} \def\yy{y}
\def\JJ{J} \def\JJr{J_{\rm (e)r}} \def\JJt{J_{\rm (e)\theta}}
\def\rr{r} \def\tt{\theta}  \def\p{P} \def\x{X}
\def\hp{\hat{P}} \def\hx{\hat{X}}

\def\af{\zeta} 
\def\RS{R^2} \def\DD{\Delta} \def\GG{G}

\def\dr{\delta r} \def\dt{\delta \theta}

\graphicspath{{pic/}}

\begin{document}
\title{Dynamics of current-carrying string loops in the Kerr naked-singularity and black-hole spacetimes}
\author{M. Kolo\v{s}}
\author{Z. Stuchl\'{\i}k}
\affiliation{Institute of Physics, Faculty of Philosophy \& Science, Silesian University in Opava, Bezru\v{c}ovo n\'{a}m.13, CZ-74601 Opava, CzechRepublic}

\begin{abstract}
Current-carrying string-loop dynamics is studied in the Kerr spacetimes. With attention concentrated to the axisymmetric motion of string loops around symmetry axis of both black-hole (BH) and naked singularity (NS) spacetimes, it is shown that the resulting motion is governed by the presence of an outer tension barrier and an inner angular momentum barrier that are influenced by the BH or NS spin. 
We classify the string  dynamics according to properties of the energy boundary function (effective potential) for the string loop motion. We have found that for NS there exist new types of energy boundary function, namely those with off-equatorial minima.
Conversion of the energy of the string oscillations to the energy of the linear translational motion has been studied. Such a transmutation effect is much more efficient in the NS spacetimes because lack of the event horizon. For BH spacetimes efficiency of the transmutation effect is only weakly spin dependent.
Transition from regular to chaotic regime of the string-loop dynamics is examined and used for explanation of the string-loop motion focusing problem. Radial and vertical frequencies of small oscillations of string loops near minima of the effective potential in the equatorial plane are given. These can be related to high-frequency quasi-periodic oscillations observed near black holes.
\end{abstract}

 
\maketitle


\section{Introduction}\label{intro}

Studies of relativistic current-carrying string loops moving axisymmetrically along the symmetry axis of Kerr or \Schw\nnd\dS{} black holes appeared recently \cite{Jac-Sot:2009:PHYSR4:,Kol-Stu:2010:PHYSR4:}. Tension of such string loops prevents their expansion beyond some radius, while their worldsheet current introduces an angular momentum barrier preventing them from collapsing into the black hole. Such a configuration was also studied in \citep{Lar:1994:CLAQG:,Fro-Lar:1999:CLAQG:}.  There is an important possible astrophysical relevance of the current-carrying string  loops \cite{Jac-Sot:2009:PHYSR4:} as they could in a simplified way represent plasma that exhibits associated string-like behavior via dynamics of the magnetic field lines in the plasma \cite{Chri-Hin:1999:PhRvD:,Sem-Dya-Pun:2004:Sci:} or due to thin isolated flux tubes of plasma that could be described by an one-dimensional string \cite{Spr:1981:AA:,Sem-Ber:1990:ASS:,Cre-Stu:2013:PhRvE:}. Motion of electrically charged string loops in combined external gravitational and electromagnetic fields has been recently studied for a Schwarzschild black hole immersed in a homogeneous magnetic field \cite{Arm-etal:2013:PRD:}.

It has been proposed in \citep{Jac-Sot:2009:PHYSR4:} that the current-carrying string loop configuration can be used as a model for jet formation - the chaotic character of the string loop motion enables strong transmutation effect when an oscillatory string-loop  motion is transformed into its translational motion with relativistic velocity $v \sim c$, even in the Schwarzschild spacetimes \cite{Stu-Kol:2012:PHYSR4:}. In the \Schw\nnd\dS{} spacetimes, the string loop translational motion is accelerated by the cosmic repulsion behind the so called static radius where the gravitational attraction of the black hole is balanced by the cosmic repulsion \cite{Stu:1983:BULAI:,Stu-Hle:1999:PHYSR4:,Stu-Schee:2011:JCAP:}, and the motion reaches the velocity of light at the cosmological horizon of the \Schw\nnd\dS{} spacetimes \cite{Kol-Stu:2010:PHYSR4:,Stu-Kol:2012:PHYSR4:}.

Recently, an extensive interest appeared in the research devoted to the Kerr superspinars, hypothetical objects indicated by String Theory \cite{Gim-Hor:2009:PhysLett:}. Exterior of a Kerr superspinar is assumed to be described by the Kerr naked singularity spacetime, while its interior is assumed to be determined by an appropriate solution of String Theory, removing the physical singularity and the causality violating region of the Kerr spacetime. The boundary of the superspinar interior is expected to be located at small radius allowing thus for appearance of all the extraordinary physical phenomena of Kerr naked singularity spacetimes \cite{deF:1974:Nat:,Stu:1980:BULAI:} related to the regions close to the physical singularity at $r=0, \theta=\pi/2$.  There are extended studies of accretion phenomena \cite{Cal-Nob:1979:NuovoCimB:,Stu:1981:BULAI:,Tak-Har:2010:CLAQG:,Bam-etal:2010:PHYSR4:,Stu-Hle-Tru:2011:CLAQG:,Stu-Sche:2012b:CLAQG:}, instabilities of test physical fields \cite{Pani-etal:2010:PHYSR4:,Dot-etal:2012:CLAQG:}, optical phenomena \cite{Stu:1980:BULAI:,Hio-Mae:2010:PHYSR4:,Bam-Fri:2009:PHYSR4:,Stu-Sche:2010:CLAQG:,Stu-Sche:2012a:CLAQG:,Sche-Stu:2013:JCAP:} and ultra-high-energy collisions \cite{Stu-Hle-Tru:2011:CLAQG:,Pat-Jos:2011:CLAQG:,Pat-Jos:2011:PHYSR4:,Stu-Sche:2012b:CLAQG:,Stu-Sche:2013:CLAQG:} in the superspinning Kerr geometry. All these studies are based on test particle (photon) or test perfect fluid approach. Therefore, it is important to study properties of the string-loop dynamics since its chaotic nature exhibits some new interesting phenomena characteristic for Kerr naked singularity spacetimes that are not present in dynamics of string loops in the field of Kerr black holes and can be astrophysically relevant.

Here, we investigate dynamics of current-carrying string loops in the field of Kerr naked singularities, generalizing thus the previous work \citep{Jac-Sot:2009:PHYSR4:} related to the string loops moving in the field of Kerr black holes. We extend also the study of the string loop motion in the Kerr black hole backgrounds and compare the naked singularity and black hole cases. Considering capture, trapping and scattering of circular string loops we identify the role of the spin parameter of the spacetime in the string dynamics. It is well known that even such a very simple axisymmetric two-body problem demonstrates apparent chaotic behavior reflected by the so called strange repeller due to the presence of the tension terms in the motion equations \citep{Fro-Lar:1999:CLAQG:}. In order to reflect the chaotic character of the string motion, we use the method of Poincare surfaces and determine the basin-boundary separating the space of initial condition due to different asymptotic outcomes of the string-loop motion. We focus attention on the transmutation effect transforming the oscillatory internal energy of the string loops into the kinetic energy of their translational motion \citep{Jac-Sot:2009:PHYSR4:,Stu-Kol:2012:PHYSR4:}, and on properties of the oscillatory motion in vicinity of potential minima where transition from regular to chaotic character of the string loop dynamics occurs.

We study a string loop threaded on to the symmetry axis of the Kerr naked singularity (black hole) chosen to be the $y$-axis - see Fig. 1. in \cite{Kol-Stu:2010:PHYSR4:}. The string loop can oscillate, changing its radius in the $x$-$z$ plane, while propagating in the $y$ direction. The string loop tension and worldsheet current form barriers governing its dynamics. These barriers are modified by the gravitational field of the Ker naked singularity (black hole)  characterized by its mass $M$ and dimensionless spin $a$. Since the mass parameter determines the distance scale of the spacetime, we focus our attention on the role of the dimensionless spin $a$ in the string loop motion. 

\section{Kerr spacetimes}

Kerr naked singularities (or exterior of Kerr superspinars) and Kerr black holes are described by the Kerr geometry that is given in the standard Boyer-Lindquist coordinates and the geometric units ($c=G=1$) in the form
\bea
 \d s^2 &=& - \left( 1- \frac{2Mr}{\RS} \right) \d t^2 - \frac{4Mra \sin^2\theta}{\RS} \, \d t \d \phi \nonumber\\
 && + \left( r^2 +a^2 + \frac{2Mra^2}{\RS} \sin^2\theta \right) \sin^2\theta \, \d \phi^2 \nonumber \\
 && + \frac{\RS}{\DD} \, \d r^2 + \RS\, \d\theta^2, 
 \label{KerrMetric} 
\eea
where
\beq
\RS = r^2 + a^2 \cos^2\theta, \quad \DD = r^2 - 2Mr + a^2,
\eeq
$a$ denotes spin and $M$ mass of the spacetimes that fulfill condition $a>M$ in the naked singularity case. The physical singularity is located at the ring $r=0, \theta = \pi/2$ that can be well characterized in the so called Kerr-Schild "Cartesian" coordinates that are related to the Boyer-Lindquist coordinated by the relations  
\bea
x &=& (r^2+a^2)^{1/2}\sin \theta\cos\left[\phi-\tan^{-1}\left(\frac{a}{r}\right)\right],\\
z &=& (r^2+a^2)^{1/2}\sin \theta\sin\left[\phi-\tan^{-1}\left(\frac{a}{r}\right)\right],\\
y &=& r\cos\theta .
\eea
Contrary to the usual notation, we change axis $z \leftrightarrow y$ in accordance with the previous works \cite{Jac-Sot:2009:PHYSR4:,Kol-Stu:2010:PHYSR4:,Stu-Kol:2012:PHYSR4:,Stu-Kol:2012:JCAP:}.
Because of the string loop axial symmetry we are interested only in constant $\phi$ sections of the whole spacetime; we are free to choose 
\beq
\phi = \tan^{-1}\left(a/r\right)
\eeq
obtaining coordinate transformation in $r,\theta$ ($x,y$) plane
\beq
 x = \sqrt{r^2 + a^2} \sin \theta ,\quad y = r \cos \theta. \label{xycord}
\eeq 
At the $x$--$y$ plane, the physical singularity is located at $x=\pm a$ and $y=0$.

In the following, we put $M=1$, i.e., we use dimensionless radial coordinate $r$ and dimensionless spin $a$. There is no event horizon in the naked singularity spacetimes, in contrast to the Kerr black hole spacetimes (with $a<1$) when two event horizons exist. The causality violation region, where $g_{\phi \phi}(r,\theta) < 0$, is determined by the condition 
\beq
       (r^2 + a^2) (r^2 + a^2 \cos^{2}\theta) + 2r a^{2} \sin^{2}\theta < 0 ,
\eeq
and it is clear immediately that this region is located at some part of the region of negatively valued radial coordinate, i.e. in the region where we ascribe negative (repulsive) gravitational mass to the Kerr spacetime \cite{Car:1973:BlaHol:}. For Kerr superspinars the extremal surface radius guaranteeing removing of the both causality violating region and the ring singularity reads $\RS(\theta)=0$. We consider this extremally limited radius in the following.

%
%
%
%

\section{Relativistic current-carrying strings loop in axially symmetric spacetimes}

The relativistic description of string motion can be given in terms of a properly chosen action reflecting both the string and spacetime properties and enabling derivation of the equations of motion. We summarize the equations of the string motion in the standard form discussed in \cite{Jac-Sot:2009:PHYSR4:}. 

The string worldsheet is described by the spacetime coordinates $X^{\alpha}(\sigma^{a})$ with $\alpha = 0,1,2,3$ given as functions of two worldsheet coordinates $\sigma^{a}$ with $a = 0,1$ that imply induced metric on the worldsheet in the form
\beq
      h_{ab}= g_{\alpha\beta}X^\alpha_{\der a}X^\beta_{\der b},
\eeq 
where $\Box_{\der a} = \partial \Box /\partial a$. The string current localized on the worldsheet is described by a scalar field $\phi({\sigma^a})$. Dynamics of the string, inspired by an effective description of superconducting strings representing topological defects occurring in the theory with multiple scalar fields undergoing spontaneous symmetry breaking \cite{Wit:1985:NuclPhysB:,Vil-She:1994:CSTD:}, is described by the action $S$ with Lagrangian $\mathcal{L}$
\beq
 S = \int \mathcal{L} \, \dif \sigma \dif \tau, \quad 
 \mathcal{L} = -(\tens + h^{ab} \varphi_{\der a}\varphi_{\der b})\sqrt{-h}.\label{katolicka_akce}
\eeq
where $ \varphi_{,a} = j_a $ determines current of the string and $\tens > 0$ reflects the string tension.

Varying the action with respect to the induced metric $h_{ab}$ yields the worldsheet stress-energy tensor density (being of density weight one with respect to worldsheet coordinate transformations)
\beq
\SS^{ab}= \sqrt{-h} \left( 2 j^a j^b - (\tens + j^2) h^{ab}\right),
\eeq
where
\beq
   j^a = h^{ab}j_{b} , \quad j^2 = h^{ab}j_{a}j_{b}.
\eeq
The contribution from the string tension with $\mu > 0$ has a positive energy density and a negative pressure (tension). The current contribution is traceless, due to the conformal invariance of the action - it can be considered as a $1+1$ dimensional massless radiation fluid with positive energy density and equal pressure \cite{Jac-Sot:2009:PHYSR4:}. 

Any two-dimensional metric is conformally flat metric, i.e., we can write
\beq
            h_{ab} = \Omega^2 \eta_{ab},
\eeq
where $\eta_{ab}$ is the flat metric and $\Omega$ is a worldsheet scalar function. Adopting coordinates $\sigma^a = (\tau, \sigma)$ such that $\eta_{\tau \sigma} = 0$ and $\eta_{\tau \tau} = -\eta_{\sigma \sigma} = -1$, the conformally flat gauge is equivalent to the conditions
\beq
    h_{\tau \sigma} = 0, \, h_{\tau \tau} + h_{\sigma \sigma} = 0, \, \sqrt{-h} h^{ab} = \eta^{ab}. \label{gCondition}
\eeq
Then the conformal factor is given by
\beq
           h_{\sigma \sigma} = \Omega^2 \eta_{\sigma \sigma} = g_{\phi \phi}.
\eeq
In the conformal gauge, the equation of motion of the scalar field reads
\beq
    \varphi_{\der \tau \tau} - \varphi_{\der \sigma \sigma} = 0.
\eeq
The assumption of axisymmetry implies that the current is independent of $\sigma$ and $j_{a,\sigma} = 0$. Using the scalar field equation of motion we can conclude that the scalar field can be expressed in a linear form with constants $j_{\sigma}$ and $j_{\tau}$
\beq
    \varphi = j_{\sigma}\sigma + j_{\tau}\tau.
\eeq
Introducing new variables 
\beq
     J^2 \equiv j_\sigma^2 + j_\tau^2, \quad \omega \equiv -j_\sigma / j_\tau ,
\eeq
we express the components of the worldsheet stress-energy density $\SS^{ab}$ in the form
\bea
&& \SS^{\tau\tau} = \frac{J^2}{g_{\phi\phi}} + \tens , \quad \SS^{\sigma\sigma} = \frac{J^2}{g_{\phi\phi}} - \tens , \label{sigma1}\\ 
&& \SS^{\sigma\tau} =  \frac{-2 j_\tau j_\sigma}{g_{\phi\phi}} = \frac{2 \omega J^2}{g_{\phi\phi} (1+\omega^2)}. \label{sigma2}
\eea
The string dynamics depends on the current through the worldsheet stress-energy tensor. The dependence is expressed using the parameters $J^2$ and $\omega$. The minus sign in the definition of $\omega$ is chosen in order to obtain correspondence of positive angular momentum and positive $\omega$.
Due to $J\leftrightarrow -J$ symmetry in Eqs (\ref{sigma1}-\ref{sigma2}), we will use only positive $J$ in following calculations, while we use three significant values $-1,0,1$ for the parameter $\omega$ characterizing the whole interval $\omega\in\langle-1,1\rangle$.

\subsection{Hamiltonian formulation of string loop dynamics}

In the Kerr spacetime we can introduce the locally non-rotating frames corresponding to zero-angular-momentum observers (ZAMO), that are co-moving with the spacetime rotation \cite{Bar:1973:BlaHol:}. ZAMO are observing the string loop in coordinates
\beq
 X^\alpha(\tau,\sigma) = (t(\tau),r(\tau),\theta(\tau),\sigma + f(\tau)). \label{strcoord}
\eeq
Now it is clear that relative to ZAMO the string loops do not rotate, and for the string coordinates (\ref{strcoord}) we can obtain the relations 
\bea
 \dot{X}^\alpha &=& X^\alpha_{\der\tau} = (t_{\der\tau}\,,r_{\der\tau}\,,\theta_{\der\tau}\,,f_{\der\tau}\,), \\
 {X'}^\alpha &=& X^\alpha_{\der\sigma} = (0,0,0,1).
\eea
For the function $f(\tau)$, we obtain
\beq
 f_{\der\tau} = -(g_{t \phi}/g_{\phi\phi}) t_{\der\tau}. 
\eeq

Varying the action with respect to $X^\mu$ implies equations of motion in the form
\beq
 \frac{\rm D}{\d \tau} P^{(\tau)}_\mu + \frac{\rm D}{\d \sigma} P^{(\sigma)}_\mu = 0, \label{EQmotion}
\eeq
where the string loop momenta are defined by the relations
\bea
 P^{(\tau)}_\mu &\equiv& \frac{\partial \mathcal{L}}{\partial \dot{X}^\mu} = \SS^{\tau a} g_{\mu \lambda} X^\lambda_{\der a} \\
 P^{(\sigma)}_\mu &\equiv& \frac{\partial \mathcal{L}}{\partial {X'}^\mu} = \SS^{\sigma a} g_{\mu \lambda} X^\lambda_{\der a}.
\eea
The equations of the motion (\ref{EQmotion}) can be expressed in the form
\beq
 {\left( \SS^{ab} g_{\mu \lambda} X^\lambda_{\der a} \right)}_{\der b} - \Gamma^\alpha_{\mu\beta} \SS^{ab} g_{\alpha \lambda} X^\lambda_{\der a} X^\beta_{\der b} = 0, \label{EQMgamma}
\eeq
where $a,b$ are coordinates $\tau,\sigma$. Using identities 
\beq
g^{\beta\lambda} \Gamma^\kappa_{\mu\beta} P_\kappa P_\lambda = \frac{1}{2} g_{\alpha\beta \der \mu} P^\alpha P^\beta
 = - \frac{1}{2} {g^{\kappa\lambda}}_{\der \mu} P_\kappa P_\lambda, 
\eeq
the equations of motion (\ref{EQMgamma}) can be rewritten in the form (6) presented in \cite{Jac-Sot:2009:PHYSR4:}.

We can define affine parameter $\af$, related to the worldsheet coordinate $\tau$ by the transformation  
\beq
 \d \tau = {\SS^{\tau\tau}} \d \af.
\eeq
Then the equations of motion (\ref{EQMgamma}) take the form
\bea
 \frac{ \d P_\mu }{\d \af} &=& - \frac{1}{2} {g^{\alpha\beta}}_{\der\mu} P_\alpha P_\beta \nn \\ 
 && - \frac{1}{2} \left[ g_{\phi\phi} (\Sigma^{\tau\tau})^2 \right]_{\der\mu}  + \frac{1}{2} \left[ g_{\phi\phi} (\SS^{\tau\sigma})^2 \right]_{\der\mu} \label{Heq1},
\eea
with the four-momentum $P_\mu$ expressed by the relation
\beq
 P_\mu \equiv P^{(\tau)}_\mu =  \SS^{\tau\tau} g_{\mu\lambda} \dot{X}^\lambda + \SS^{\tau\sigma}g_{\mu\lambda} {X'}^\lambda. \label{Heq2}
\eeq

Considering the Hamilton equations
\beq
 \frac{\d \x^\mu}{\d \af} = \frac{\partial H}{\partial \p_\mu}, \quad
 \frac{\d \p_\mu}{\d \af} = - \frac{\partial H}{\partial \x^\mu}, \label{Ham_eq}
\eeq
the equations (\ref{Heq1}-\ref{Heq2}) imply the Hamiltonian in the form
\beq
 H = \frac{1}{2} g^{\alpha\beta} \p_\alpha \p_\beta + \frac{1}{2} g_{\phi\phi} \left[(\SS^{\tau\tau})^2 - (\SS^{\tau\sigma})^2 \right] \label{AllHam}
\eeq
where $\alpha, \beta$ are the spacetime coordinates $ t,r,\theta,\phi$. 
 
\subsection{Integrals of the motion}

The Kerr metric (\ref{KerrMetric}) does not depend on coordinates $t$ (stacionarity) and $\phi$ (axial symmetry). Such symmetries  imply conserved quantities - string energy and axial angular momentum. The string energy $E$ reads
\bea
 - E &=& P_t = g_{tt} \SS^{\tau\tau} X^t_{\der\tau} + g_{t\phi} (\SS^{\tau\tau} X^\phi_{\der\tau} + \SS^{\tau\sigma} X^\phi_{\der\sigma}) \nonumber \\ 
 &=& \SS^{\tau\tau} \left(g_{tt} -g_{t\phi}^2/g_{\phi\phi} \right) t_{\der\tau} + g_{t\phi} \SS^{\sigma\tau}. \label{Cenergy}
\eea
The string loops do not rotate, but have nonzero angular momentum, completely generated by the current on the string. The axial component of the angular momentum reads
\bea
 L &=& P_\phi = g_{\phi t} \SS^{\tau\tau} X^t_{\der\tau} + g_{\phi \phi} (  \SS^{\tau\tau} X^\phi_{\der\tau} + \SS^{\tau\sigma} X^\phi_{\der\sigma} ) \nonumber \\
 	 &=& g_{\phi\phi} \SS^{\sigma\tau} = -2 j_\tau j_\sigma. \label{Cmoment}
\eea
From equations (\ref{sigma1}-\ref{sigma2}) we clearly see that $\omega=-1$ represents string loops with negative axial angular momentum $L<0$, $\omega=~0$ represents the string loops with $L=0$, and $\omega=+1$ represents the string loops with $L>0$. The string loops having no scalar field $\varphi$ feel no centrifugal barrier because their angular momentum parameter $J=0$ (Nambu---Goto strings) - naturally, they have  also $L=0$; such situation is different from $\omega=0, J\neq0$ case, where the centrifugal repulsion governed by the non zero $J$ parameter is present. Recall that in the non-rotating Schwarzschild spacetime, the string loop equations of motion are independent of the parameter $L$ \cite{Kol-Stu:2010:PHYSR4:}. 

The string dynamics depends on the current $J$ through the worldsheet stress-energy tensor. Using the two constants of motion (\ref{Cenergy}-\ref{Cmoment}), we can rewrite the Hamiltonian (\ref{AllHam}) into the form related to the $r$ and $\theta$ momentum components
\bea
 H &=& \frac{1}{2} g^{rr} \p_r^2 + \frac{1}{2} g^{\theta\theta} \p_\theta^2 \nonumber\\
 && + \frac{1}{2} g_{\phi\phi} (\SS^{\tau\tau})^2 + \frac{g_{\phi\phi}(E + g_{t\phi} \SS^{\sigma\tau})^2}{2 (g_{tt} g_{\phi\phi}-g_{t\phi}^2)}. \label{HamHam}
\eea
The equations of motion (\ref{Ham_eq}) following from the Hamiltonian (\ref{HamHam}) are very complicated and can be solved only numerically in general case, although there exist analytical solutions for simple cases of the motion in the flat or de~Sitter spacetimes \cite{Kol-Stu:2010:PHYSR4:}.

%
%
%
%
\begin{figure*}[htp]
\subfigure[ \,\, $J=0, \,\, a=0$]{\includegraphics[width=0.31\hsize]{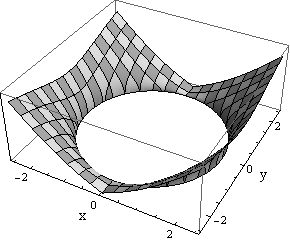}}
\subfigure[ \,\, $J=0, \,\, a=0.99$]{\includegraphics[width=0.31\hsize]{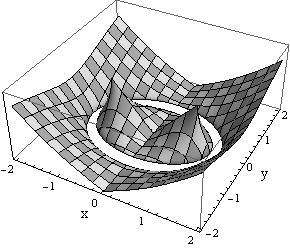}}
\subfigure[ \,\, $J=0, \,\, a=1.1$]{\includegraphics[width=0.31\hsize]{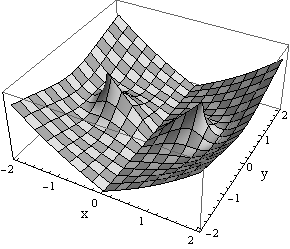}}
\subfigure[ \,\, $J=0.1, \,\, a=0$]{\includegraphics[width=0.31\hsize]{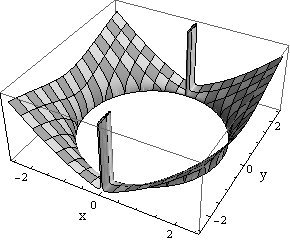}}
\subfigure[ \,\, $J=0.1, \,\, a=0.99$]{\includegraphics[width=0.31\hsize]{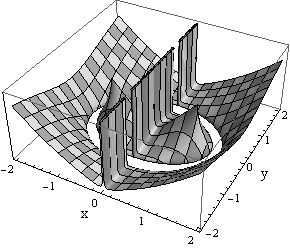}}
\subfigure[ \,\, $J=0.1, \,\, a=1.1$]{\includegraphics[width=0.31\hsize]{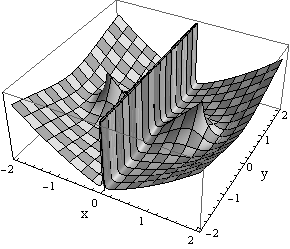}}
\subfigure[ \,\, $J=1.1, \,\, a=0$]{\includegraphics[width=0.31\hsize]{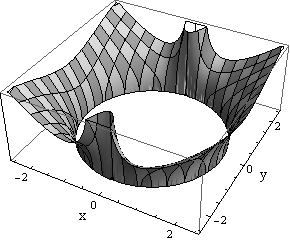}}
\subfigure[ \,\, $J=1.1, \,\, a=0.99$]{\includegraphics[width=0.31\hsize]{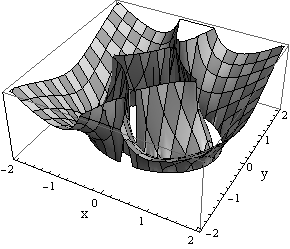}}
\subfigure[ \,\, $J=1.1, \,\, a=1.1$]{\includegraphics[width=0.31\hsize]{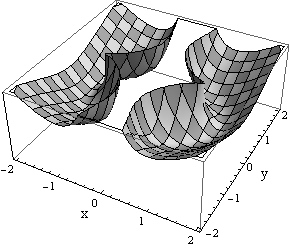}}
\caption{Energy boundary function $ E_{\rm b}(x,y;J,\omega,a)$ for string loops with $J=0$ (a-c) and $J\neq 0$ (d-i), given for characteristic values of spin $a$. We chose parameter $\omega=0$ in all the presented cases, but the function $ E_{\rm b}(x,y;J,\omega)$ depends only slightly on $\omega$ for small values of $J$ and its behavior is of the same character for the other values of $\omega$. The maxima ("horns") are located at the singularity points $[\pm a,0]$. Empty space represents the dynamical region (\ref{dynreg}), where $ E_{\rm b}$ is not defined. For angular momenta small enough  ($J=0.1$), the repulsive centrifugal barrier is located at the ring $x~\leq~a$, while for angular momentum large enough ($J=1.1$), it is located above the ring singularity.
\label{plotJ0tri}}
\end{figure*}

\section{Energy boundary function}

It is useful to examine regions allowed for the motion of string loops according to their energy. The loci where the string loops have zero velocity ($\dot{r}=0, \dot{\theta}=0$) form boundary of the string-loop motion. The boundary energy function can be defined by the relation 
\beq
 E = E_{\rm{b}}(r,\theta) = \sqrt{g_{t\phi}^2-g_{tt}g_{\phi\phi}} \, \SS^{\tau\tau} - g_{t\phi}\SS^{\sigma\tau} \label{EqEbRT} \label{StringEnergy},
\eeq  
implied by the condition $H=0$, given by reparameterization invariance of the action (\ref{katolicka_akce}).

We can make the rescaling $E_{\rm b} \rightarrow E_{\rm b} / \mu $ and $J \rightarrow J / \sqrt{\mu} $,  assuming $\mu > 0$. This choice of ``units'' will not affect energy boundary function and is equivalent to setting the string tension $\mu=1$ in Eqs (\ref{sigma1}-\ref{sigma2}), (\ref{EqEbRT}).

The energy boundary function then, in the Boyer-Lindquist $r,\theta$ coordinates, takes the form  
\beq
 E_{\rm b} (r,\theta,J,\omega) = \frac{4 a \omega J^2 r }{\left(\omega^2+1\right) \GG }+\sqrt{\Delta}
   \left(\frac{J^2 R^2}{\GG \sin(\theta)}+\sin(\theta)\right), \label{EBrtKerr}
\eeq
where we introduced the function
\beq    
  \GG = \left(a^2+r^2\right) R^2 +2 a^2 r \sin^2(\theta ).
\eeq
Clearly, the energy boundary function (\ref{EBrtKerr}) is not defined in the so called dynamical region, located between the black hole horizons, where 
\beq
g_{t\phi}^2-g_{tt}g_{\phi\phi} = \DD \sin^2 (\theta) < 0. \label{dynreg}
\eeq

The critical points of the boundary energy function are located where the function $E_{\rm b}(r,\theta;J,\omega,a)$ is not differentiable or its derivative is zero (stationary points). The stationary points are determined by the stationarity conditions
\bea
  (E_{\rm b})_{r}' &=& 0 \label{extr_a1} \\
  (E_{\rm b})_{\theta}' &=& 0  \label{extr_a2}
\eea
The notation $()_{m}'$ corresponds to the derivation with respect to the coordinate $m$. In order to determine character of the stationary points denoted as $(r_{\rm e},\theta_{\rm e})$, we have to consider the stability conditions, giving a maximum or a minimum of the energy boundary function $E_{\rm b}(r,\theta;J,\omega,a)$
\bea
 && [(E_{\rm b})_{rr}''] (r_{\rm e},\theta_{\rm e}) \quad < 0 \,\, \mathrm{(max)} \quad > 0 \,\, \mathrm{(min)}  \label{extr_b1}\\
 && [(E_{\rm b})_{rr}'' (E_{\rm b})_{\theta\theta}'' - (E_{\rm b})_{r\theta}'' (E_{\rm b})_{\theta r}''](r_{\rm e},\theta_{\rm e}) > 0 . \label{extr_b2}
\eea
Behavior of the energy boundary function $E_{\rm b}(r,\theta;J,\omega,a)$ in each of the directions $r$ and $\theta$ is relevant for the character of the string-loop motion boundary. We first discuss behavior of $E_{\rm b}(r,\theta;J,\omega,a)$ for the Nambu---Goto strings with $J=0$ and then for current-carrying strings with $J\neq0$.

\subsection{Nambu–--Goto string loops}

\begin{figure}[htp]
\subfigure[ \quad BH $a=0.99,\, J=0$]{\includegraphics[width=0.48\hsize]{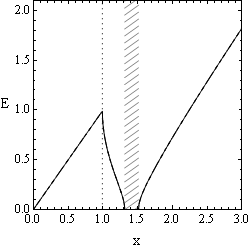}}
\subfigure[ \quad BH $a=0.99,\, J=2$]{\includegraphics[width=0.48\hsize]{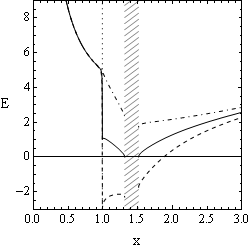}}
\subfigure[ \quad NS $a=1.1,\, J=0$]{\includegraphics[width=0.48\hsize]{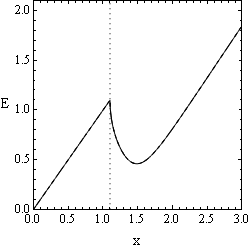}}
\subfigure[ \quad NS $a=1.1,\, J=2$]{\includegraphics[width=0.48\hsize]{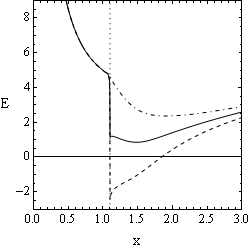}}
\caption{Equatorial ($y=0$) sections of the energy boundary function $E_{\rm b}(x,y;J,\omega)$ for Kerr BH ($a=0.99$) and NS ($a=1.1$) in the neighborhood of the Kerr ring singularity (dotted line). Particular values of the parameter $\omega$ are distinguished as $-1$ (dashed), $0$ (plain) and $1$ (dash-dotted) curve. Dynamical region of the Kerr BH is dashed. Notice that in the case $\omega=-1$,  corresponding to axial angular momentum $L<0$, negative energies $E<0$ are allowed. \label{sectionEb}}
\end{figure}

We know that due to the transmutation effect, the string-loop maximal acceleration in the $y$-direction is obtained for small currents $\JJ \sim 0$ \cite{Stu-Kol:2012:PHYSR4:}. It it obvious to examine the  motion for the most simplified situation - relativistic string loops with no current $J=0$. 

Energy boundary function $ E_{\rm b}(r,\theta;a)$ then reduces to the form
\beq
 E^2_{\rm b} = g_{t\phi}^2 - g_{tt}g_{\phi\phi} = \sin^2(\theta) (a^2 - 2r + r^2). \label{eqEbJ0}
\eeq
The function $ E_{\rm b}(x,y;a)$ is plotted for parameters $a\in\{0,0.99,1.1\}$ in Fig. \ref{plotJ0tri}(a-c); in these 3D plots we see overall behavior of the function $ E_{\rm b}(x,y;a)$ near the origin of coordinates. All three cases merge at large distances. Here we use the Cartesian (Kerr-Schild) coordinates (\ref{xycord}) that are convenient to reflect properly the Kerr spacetime near the physical ring singularity.

The critical points of $E_{\rm b}(x,y;a)$ are located at 
\bea
&(A)& \, \text{maxima:} \,\, A=[\pm a,0]; \quad E_{\rm b} (A) = a \\
&(B)& \, \text{saddle:} \quad B=[\pm \sqrt{1+a^2},0]; \nonumber \\ 
&& \qquad \qquad     E_{\rm b} (B) = \sqrt{a^2-1} \\
&(C)& \, \text{valley:} \quad x = 0; \quad E_{\rm b}(0,y) = 0.
\eea
Point $A$ (top of the horn) is physically forbidden position in the Kerr spacetimes as it corresponds to the Kerr ring singularity, but it can be realized in the field of properly constructed Kerr superspinars (with $R=0$), point $B$ (pass between the horn and the wall) exists only for Kerr naked singularities. String loops located at points $C$ (valley between horns) degenerate to have zero radius. At all three points ($A,B,C$) the string can be at rest; points $A$ an $B$ are unstable against perturbations and point $C$ is stable against perturbations in the $x$-direction only. 

For Nambu---Goto string loops no stable equilibrium positions exist (minima in $E_{\rm b}(r,\theta;a)$ function) -- this is caused by vanishing of the centrifugal barrier due to $J=0$.

\subsection{Current-carrying string loops}

In the case $J=0$, we have demonstrated pure interplay of the influence of gravity and string tension in the behavior of the energy boundary function $E_{\rm b}(x,y;a)$. For the flat spacetime and nonzero angular momentum $J$, the energy boundary function $E_{\rm b}(x,y;J)$ takes the form 
\beq
  E_{\rm b}(x,y) = x + \frac{J^2}{x}, \label{EBflat}
\eeq
representing long "valley" along $x=J$ -- see \cite{Kol-Stu:2010:PHYSR4:}. In the case of Kerr geometry, with arbitrary parameters $a,J,\omega$, the $E_{\rm b}(x,y;J,\omega,a)$ function demonstrates a complicated form that can be imagined as arising due to mixing of the effects of the string tension and angular momentum $J$ induced by the string current, as in flat spacetime (\ref{EBflat}), and the influence of the Kerr spacetime (gravitation) for $J=0$ (\ref{eqEbJ0}). Combination of both effects  leads to a very complex behavior of $E_{\rm b}(x,y;J,\omega,a)$ especially in the region near the  physical ring singularity and becomes to be strongly dependent on the parameter $J$ governing the centrifugal repulsive barrier, and on the spin in the case of the Kerr NS spacetimes.

\begin{figure}
\subfigure[\quad BH $a=0.99$]{\includegraphics[width=0.49\hsize]{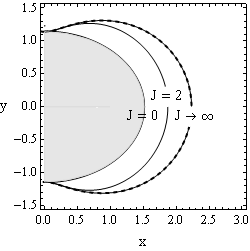}}
\subfigure[\quad NS $a=1.1$]{\includegraphics[width=0.49\hsize]{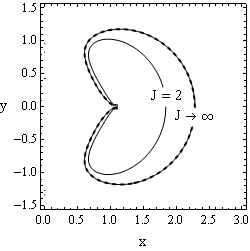}}
\caption{
String loops with negative energy ($E<0$) can exist only inside the ergosphere, and for string loops with $\omega=-1$. Contours $E_{\rm b}(x,y)=0$ with typical values of the parameter $J$ (solid curves) are plotted; for $J=0$ they coincide with outer the horizon, for $J\rightarrow\infty$ they coincide with the boundary of the ergosphere, so called surface of static limit (thick dotted).
\label{negENfig}}
\end{figure}

We express the current-carrying string loop ($J>0$) energy boundary function $E_{\rm{b}}$ also in the Kerr---Schild coordinates $x,y$ (\ref{xycord}), when it takes the form
\beq
 E_{\rm{b}}(x,y) = \Big\{ { \quad x + \JJ^2/x  \quad \textrm{for} \quad x \leq a, y = 0 \atop
 E_{\rm{b}}[r(x,y),\theta(x,y)] \quad {\rm elsewhere};} \label{EExyDef}
\eeq
we define the energy boundary function $E_{\rm{b}}(x,y)$ also for region $x\leq{a}, y=0$ to be continuously matched at $x=a$ in the case of $\omega = 1$. At the Kerr ring singularity $[x,y]=[a,0]$, the function is defined as $E_{\rm b}(a,0)=a+J^2/a$, but the limit of the external ($x>a$) energy boundary function $E_{\rm b}(r,\theta;J,\omega,a)$ reads
\beq
 \lim_{r \rightarrow 0} E_{\rm{b}}(r,\pi/2) = a+\frac{2 J^2 \omega}{a (1+\omega^2)}. \label{lowE}
\eeq
The only essential discontinuity exists at the ring singularity for values of parameters $J>0$ and $\omega\in\{-1,0\}$. The region $x\leq{a}, y=0$ is hidden below the outer event horizon for Kerr BH; the discontinuity is important for Kerr NS only, see Fig. \ref{sectionEb}.

Behavior of the energy boundary function near the physical ring singularity where the centrifugal barrier is relevant is demonstrated in Fig. \ref{plotJ0tri}(d-f) for small $J$ when the barrier is located at the circle $|x|<a$ (cases e,f, under the ring singularity), and in (g-i) for larger $J$ when the barrier is above the ring singularity (case h,i).

\begin{figure*}[htp]
\subfigure[ \quad $a=0$]{\includegraphics[width=0.24\hsize]{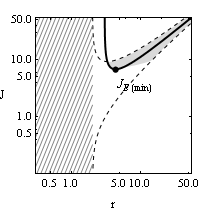}}
\subfigure[ \quad $a=0.99, \omega=-1$]{\includegraphics[width=0.24\hsize]{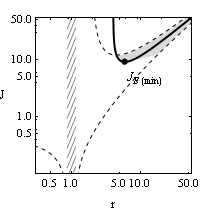}}
\subfigure[ \quad $a=0.99, \omega=0$]{\includegraphics[width=0.24\hsize]{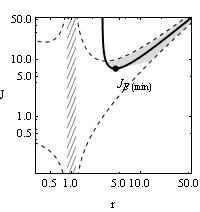}}
\subfigure[ \quad $a=0.99, \omega=+1$]{\includegraphics[width=0.24\hsize]{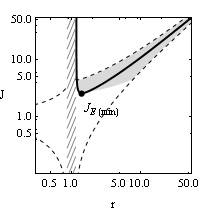}}
\subfigure[ \quad $a=1.1, \omega=-1$]{\includegraphics[width=0.24\hsize]{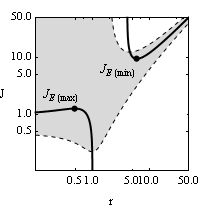}}
\subfigure[ \quad $a=1.1, \omega=0$]{\includegraphics[width=0.24\hsize]{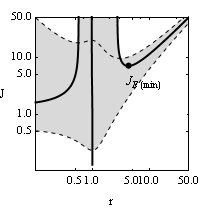}}
\subfigure[ \quad $a=1.2, \omega=0$]{\includegraphics[width=0.24\hsize]{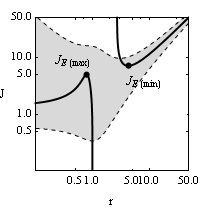}}
\subfigure[ \quad $a=1.1, \omega=+1$]{\includegraphics[width=0.24\hsize]{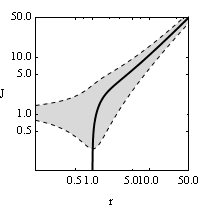}}
\caption{Function $\JJ^2_{\rm (e)eq}(r;a,\omega)$ (black thick curve) denoting the local extrema of the energy boundary function $E_{\rm b}(r,\theta;J,\omega,a)$ located in the equatorial plane. We plotted the function $\JJ^2_{\rm (e)eq}(r;a,\omega)$ for characteristic combinations of the parameters $a$ and $\omega$. The dotted curves represent the functions $J_{\rm L1}(r;a,\omega)$ and $J_{\rm L2}(r;a,\omega)$ determining the regions from which the string loops can escape to the infinity along the $y$ axis.  In the shaded regions the string loops are trapped by closed energy boundaries, given by the sections $E=E_{\rm b}(x,y;J,\omega,a)$. Note that string loops can be captured by Kerr BH, therefore,  we have to exclude such cases from the gray region; 
for Kerr NS only the trapped states (gray) can exist for all values of the parameter $J$ satisfying  condition $J_{\rm L1}<J<J_{\rm L2}$, since no capture of string loops by Kerr NS is possible.
Hatched are the dynamical regions of the Kerr spacetime.
Fig. (a) with $a=0$ represents string loops in the \Schw{} spacetime and corresponds to the Fig 8(a) presented in  \cite{Kol-Stu:2010:PHYSR4:}. \label{picJE2}
}
\end{figure*}

The energy boundary function $E_{\rm{b}}(r,\theta;J,\omega,a)$, given by Eq. (\ref{EBrtKerr}), is always positive for $\omega\in\{0,1\}$; however, for $\omega=-1$ there is a region determined by the condition 
\beq
 \sqrt{\Delta} \left(\GG \sin (\theta)^2 + J^2 R^2 \right)- 2 a r J^2 \sin(\theta) \leq 0 \label{negE}
\eeq
where string loops with negative energy $E<0$ can exist, as demonstrated in Fig. \ref{sectionEb}. 
It is useful to relate the states corresponding to the boundary energy function with $E<0$ to the ergosphere of the Kerr geometry - we can demonstrate that the string loops with negative energy can exist only inside the ergosphere, see Fig. \ref{negENfig}.

The behavior of the energy boundary function $E_{\rm}(x,y)$ far away from the origin of coordinates, where the spacetime is flat, is given by (\ref{EBflat}) -- this case was discussed in \cite{Kol-Stu:2010:PHYSR4:} and will not be repeated here.

\begin{figure*}[htp]
\subfigure[ $\quad \omega=-1$]{\includegraphics[width=5.6cm]{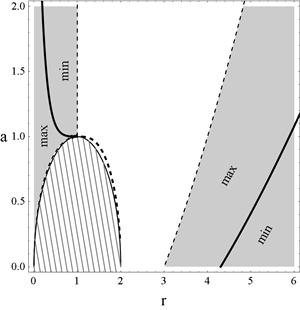}}
\subfigure[ $\quad \omega=0$]{\includegraphics[width=5.6cm]{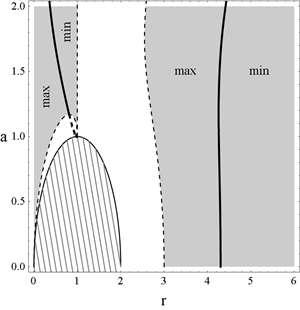}}
\subfigure[ $\quad \omega=+1$]{\includegraphics[width=5.6cm]{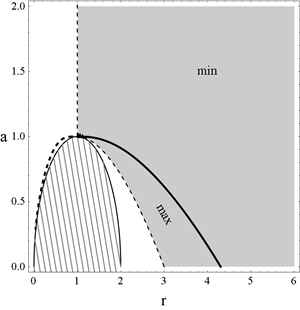}}
\caption{Loci of the local extrema of the energy boundary function $E_{\rm b}(r,\theta;J,\omega,a)$ in the equatorial plane in dependence on the rotation parameter $a$, allowing us to classify different regimes of behavior of the energy boundary function $E_{\rm b}$. 
Thin solid curves give the position of the horizons and restrict the dynamical region (hatched). Gray color denotes region where the function $J^2_{\rm (e)eq}(r;\omega,a)$ is positive - extrema of energy boundary function $E_{\rm b}(r,\theta;J,\omega,a)$ can exist in this area, while in the white areas extrema of $E_{\rm b}(r,\theta;J,\omega,a)$ cannot exist.
The regions with maxima of $E_{\rm b}$ and regions with minima of $E_{\rm b}$ are separated by thick black curve, which is given by the extrema of the $J^2_{\rm (e)eq}(r,\omega,a)$ function.
It is useful to compare this figure with Fig. \ref{picJE2}, where exact functions $J^2_{\rm (e)eq}(r;\omega,a)$ are plotted.
It can be seen that for $\omega~\in~\{-1,1\}$ only two types of behavior of $E_{\rm b}$ exist: Kerr BH and Kerr NS, while three types of this behavior occur for $\omega = 0$: Kerr BH, Kerr NSa and Kerr NSb.
\label{FIGeqplane}
}
\end{figure*}

\begin{figure}
\includegraphics[width=0.75\hsize]{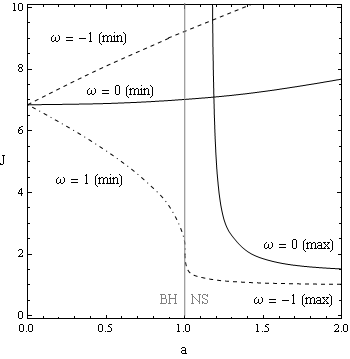}
\caption{Extrema of the $J^2_{\rm(e)eq}(r;\omega,a)$ function in dependence on the Kerr rotational parameter $a$.
Particular values of the parameter $\omega$ are distinguished as $-1$ (dashed), $0$ (plain) and $1$ (dash-dotted) curve.
\label{JEextrm}
}
\end{figure}

\section{Equilibrium positions of string loops}

In the Kerr spacetimes, the stationary points of the energy boundary function $E_{\rm}(r,\theta;J,\omega,a)$, governing the equilibrium positions of string loops, can be expressed using Eqs. (\ref{extr_a1}-\ref{extr_a2}). The stationary points can be related to the angular momentum parameter $J$, and sorted into two groups:
\begin{itemize}
\item points in {\it equatorial plane}, given by the condition 
\beq
 J^2 = J^2_{\rm (e)r}(r,\pi/2;\omega,a), \quad \tt=\pi/2
\eeq
\item points {\it off equatorial plane}, given by the condition 
\beq
 J^2 = \JJr^2(r,\theta;\omega,a), \quad J^2 = \JJt^2(r,\theta;\omega,a) . \label{offEX}
\eeq
\end{itemize}
We define the functions $\JJr^2(r,\theta,;\omega,a),\JJt^2(r,\theta;\omega,a)$ governing the stationary points (local extrema) of the energy boundary function by the relations
\begin{widetext}
\bea
\JJr^2(r,\theta) &\equiv&  \frac{ (1+\omega^2) \GG^2 (2r-2) \sin^2(\theta)}{ 8 a \sqrt{\DD} \, \omega \sin(\theta) F_1+ (1+\omega^2) ( \GG [ 2 (r-1) R^2 + 4r \DD ] - F_2)},  \label{JExR} \\
F_1 &=& \GG -4 r^4 -2 a^2 r^2-2 a^2 r \left(r \cos ^2(\theta )+\sin ^2(\theta )\right),  \\
F_2 &=& 4 R^2 \DD \left[r \left(a^2+2 r^2\right)+a^2 \left(r \cos^2(\theta)+\sin^2(\theta )\right)\right],   \\
\JJt^2(r,\theta) &\equiv& \frac{ (1+\omega^2) \GG^2 }{ (1+\omega^2) ( 2\GG a^2 + R^2 [\GG \csc^2(\theta) -2 a^2\DD]) -8 a^3 \omega \sqrt{\DD} \, r \sin(\theta) }.  \label{JExT}
\eea
\end{widetext}
In equatorial plane, the local extrema of the energy boundary function $E_{\rm b}(r,\theta;J,\omega,a)$ were previously found in both BH and NS spacetimes \cite{Jac-Sot:2009:PHYSR4:,Kol-Stu:2010:PHYSR4:,Stu-Kol:2012:JCAP:} -- here they will be discussed in detail in section \ref{ExInEqPlane}; the off-equatorial extrema were observed only in the NS spacetimes \cite{Stu-Kol:2012:JCAP:}, they will be discussed in detail in section \ref{ExOffEqPlane}.

For classification of the string loop dynamics it is crucial, if the constant-energy section of the  energy boundary function $E=E_{\rm b}(r,\theta;J,\omega,a)$ is open to infinity in the $y$-direction - the string loops can then escape along the $y$-axis. The necessary energy condition for escaping in the Kerr spacetimes is given by the relation 
\beq
 E > E_{\rm b (0)min}(r,\pi/2) = 2 \JJ, \label{minEB}
\eeq
determined by the minimal energy of string loops at infinity. (Recall that in the asymptotically \dS{} spacetimes the escaping condition is more complex \cite{Kol-Stu:2010:PHYSR4:}.) If the energy is not sufficiently high, the string loops are captured by the BH, or trapped in some "lake" like region in the BH or NS spacetimes. Using relation (\ref{minEB}), we can express condition for escaping string loops by the relation
\beq
 J_{\rm L1} < J < J_{\rm L2}
\eeq
where we define new functions $J_{\rm L1}$ (with $+$) and  $J_{\rm L2}$ (with $-$) 
\bea
 && J_{\rm L1,L2}(r) = \frac{(1+\omega^2) \GG \pm \sqrt{(1+\omega^2) \GG F_3}}{ (1+\omega^2) \csc(\theta) \sqrt{\DD} R^2 + 4 a r \omega }, \label{yeject} \\
  && F_3 = (1+\omega^2)(\GG-\DD R^2) -4 \sqrt{\DD} a r \omega \sin(\theta).
\eea
The behavior of the functions $J_{\rm L1}(r;\omega,a)$ and  $J_{\rm L2}(r;\omega,a)$ is demonstrated in Fig. \ref{picJE2}.

\begin{figure*}[htp]
\includegraphics[width=0.7\hsize]{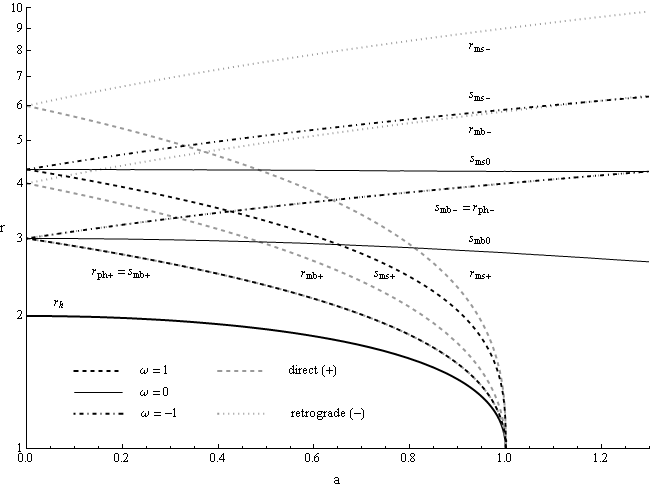}
\caption{
Characteristic radii of the string loop dynamics in the Kerr spacetimes introduced in this paper are  compared to characteristic radii of the circular equatorial orbits of test particles and photons. We plotted loci of the outer BH horizon $r_{\rm h}$, photon circular orbits $r_{\rm ph}$, innermost bound circular orbits $r_{\rm mb}$ and innermost stable circular orbits $r_{\rm ms}$; we use dashed curves for the direct (corotating) orbits $(+)$ and dotted curves for the retrograde orbits $(-)$ - compare with Fig. 1. from \cite{Bardeen:1972:AJ:}
The innermost stable positions of string loops $s_{\rm ms}$ are determined by the local minima of the  $J_{\rm (E)eq}(r;\omega,a)$ function. The innermost bound positions of string loops $s_{\rm mb}$ are given by the minimal value of the radial coordinate $r$ for which the function $J_{\rm (E)eq}(r;\omega,a)$ exist, see Figs. \ref{picJE2} and \ref{FIGeqplane}. The innermost stable and innermost bound positions are defined separately for string loops having $\omega = -1, 0, +1$. \label{IMPradii}
}
\end{figure*}

%
\subsection{Extrema in the equatorial plane} \label{ExInEqPlane}
%

For small values of $J<a$, the energy boundary function $E_{\rm b}(x,y;J,\omega,a)$, given by (\ref{EExyDef}), allows in the region $x~\leq~a$ of the equatorial plane ($y=0$) a minimum at $x=J$; moreover, there is its maximum at ring singularity $x=a$. Such points are for Kerr BH spacetimes  located behind the event horizon (\ref{dynreg}); therefore, we consider them only in the case of Kerr NS spacetimes. 

At the region above the ring singularity ($x>a$) we can examine the local extrema of the energy boundary function $E_{\rm b}(r;J,\omega,a)$ using the radial coordinate $r$ instead of the Cartesian coordinate $x$ without loss of generality. In the equatorial plane ($y=0$, $\theta = \pi/2$), equations (\ref{JExR}-\ref{JExT}) imply that the stationary points are given by the condition
\beq
 J^2 = J^2_{\rm(e)eq}(r;\omega,a) \equiv \JJr^2(r,\pi/2,\omega,a), \label{JeEQ}
\eeq
where
\bea
 && J^2_{\rm(e)eq}(r)  =  \frac{(r-1) \left(\omega^2+1\right) \left(a^2 (r+2)+r^3\right)^2}{4 a \omega \sqrt{\DD} \left(a^2+3 r^2\right)+\left(\omega^2+1\right) F_4 }, \label{eqrovina} \\
 && F_4 = (r-3) r^4 -2 a^4+a^2 r \left(r^2-3 r+6\right).
\eea
The relevant examples of the behavior of the $J^2_{\rm(e)eq}(r;\omega,a)$ function are shown in the Fig. \ref{picJE2} for characteristic values of the parameters $a,\omega$. A given string current $J$ (constant of motion) represented by a line has possible multiple intersections with the function  $J^2_{\rm(e)eq}(r)$ that determine positions of the local extrema of the $E_{\rm b}$ function, i.e., the equation (\ref{JeEQ}) governs the stationary points of the energy boundary function $E_{\rm b}$. 
Notice that in the Kerr BH case, the behavior of the $J^2_{\rm(e)eq}(r;\omega,a)$ and $J_{\rm L1}(r;\omega,a)$, $J_{\rm L2}(r;\omega,a)$ functions is of the same character for all three values of the parameter $\omega$, while significant differences occur in the Kerr NS case. 

The local extrema of the $J^2_{\rm(e)eq}(r;\omega,a)$ function, given by the condition 
$(J^2_{\rm(e)eq}\,)_{r}'~=~0$, enable us to distinguish maxima and minima of the energy boundary function $E_{\rm b}(r;J,\omega,a)$. The local extrema of the $J^2_{\rm(e)eq}(r;\omega,a)$ are given for each value of the parameter $\omega=-1,0,+1$ by a very complex equation mixing the radial coordinate $r$ and the spin $a$ and giving the function $a_{\rm extr}(r;\omega)$ separating the regions corresponding to local maxima and minima of the energy boundary function. Therefore, we do not present this equation explicitly here; the results of a numerical treatment are plotted in the $a$--$r$ plane in Fig. \ref{FIGeqplane} where the function $a_{\rm extr}(r;\omega)$ is represented by thick black curves. For completeness, we also give the functions ${J^{2 \,(\rm extr)}_{\rm(e)eq}}(a;\omega)$ determing the values of the parameter J at the extremal points $J_{\rm E(min)},J_{\rm E(max)}$ as a function of the spin parameter $a$ in Fig. \ref{JEextrm}. 

The behavior of the energy boundary function $E_{\rm b}(r;J,\omega,a)$ in the equatorial plane is completely given by the function $J^2_{\rm(e)eq}(r;\omega,a)$. We can distinguish three ($\omega\in{-1,0,1}$) types of the $\JJ^2_{\rm (e)eq}(r;\omega,a)$ function behavior for Kerr BH (see Figs \ref{picJE2}(b,c,d)), while four types are relevant for Kerr NS (see Figs \ref{picJE2}(e,f,g,h). 
The Kerr NS case with $\omega=0$ demonstrates two subcases in dependence on the spin parameter $a$: for  $1<a<a_{\rm crit}$, only one minimum of the $\JJ^2_{\rm (e)eq}(r;\omega,a)$ function exists, for  $a>a_{\rm crit}$, one minimum and one maximum of the $\JJ^2_{\rm (e)eq}(r;\omega,a)$ function exist. The critical value of the spin $a_{\rm crit}$ reads
\beq
a_{\rm crit}  \doteq 1.169052.
\eeq


At $x>a$, the string loop equilibrium position closest to the ring singularity ($x=a$) can be  determined from the local extrema of the energy boundary function $E_{\rm b}$. In the equatorial plane,  the marginally unstable (bound) equilibrium position $s_{\rm mb}(a;\omega)$ corresponds to the maximum of the energy boundary function $E_{\rm b}$, and the marginally stable equilibrium position $s_{\rm ms}(a;\omega)$ corresponds to the minimum of the $E_{\rm b}$ function. The marginally unstable and marginally stable equilibrium radii of the string loops, $s_{\rm mb}$ and $s_{\rm ms}$, can be compared to the radii of the marginally bound $r_{\rm mb}$ and marginally stable $r_{\rm ms}$ circular orbits of test particles in the Kerr spacetimes \cite{Bardeen:1972:AJ:,Stu:1980:BULAI:}. Radii of all the important orbits are plotted in Fig. \ref{IMPradii}, where we use $+$ symbol for direct particle orbits (i.e. corotating with positive angular momentum $L>0$), and $-$ for retrograde orbits (counterrotating with $L<0$).

For $\omega=\pm~1$, radii of the string loop marginally bound equilibria $s_{\rm mb\pm}$ coincide with radii of the direct or retrograde photon orbits $r_{\rm{ph}\pm}$ \cite{Mis-Tho-Whe:1973:Gra:}, while the radii of marginally stable equilibria $s_{\rm ms\pm}$ are located between the radii of marginally bound equilibria $s_{\rm mb}$ and radii of the marginally stable particle orbits $r_{\rm ms\pm}$.
Recall that the value $\omega=1$ represents positive string loop angular momentum $L>0$ (\ref{Cmoment}), while $\omega=-1$ corresponds to $L<0$, as in the case of test particles.
The case $\omega=0$, with vanishing axial angular momentum ($L=0$) has no test particle equivalent - see Fig. \ref{IMPradii}.


\begin{figure*}
\subfigure[\quad $\omega=-1$]{\includegraphics[width=0.3\hsize]{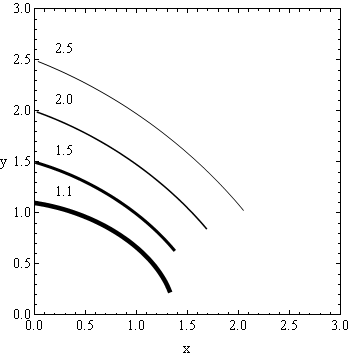}}
\subfigure[\quad $\omega=0$]{\includegraphics[width=0.3\hsize]{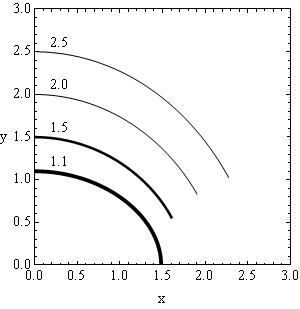}}
\subfigure[\quad $\omega=1$]{\includegraphics[width=0.3\hsize]{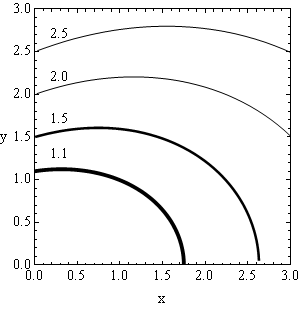}}
\caption{Positions of the off-equatorial local extrema (minima) of the energy boundary function $E_{\rm b}(x,y;J,\omega,a)$, numerically calculated from eq. (\ref{offEX2}), are presented for representative values of the rotation parameter $a>1$ (Kerr NS). For $a\leq1$ (Kerr BH) no off-equatorial local extrema can be found. 
\label{offEQex} }
\end{figure*}

\begin{figure}
\subfigure[ \quad $E=0.9, J=0.6$ ]{\includegraphics[width=0.48\hsize]{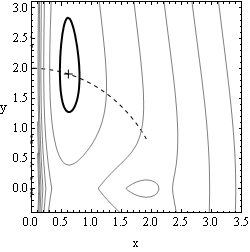}}
\subfigure[ \quad $E=1.3, J=0.6$ ]{\includegraphics[width=0.48\hsize]{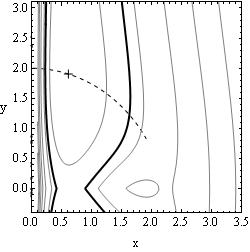}}
\subfigure[ \quad $E=1.7, J=1.2$ ]{\includegraphics[width=0.48\hsize]{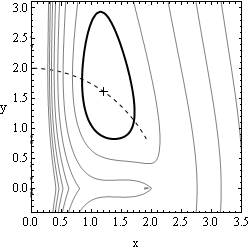}}
\subfigure[ \quad $E=2.1, J=1.6$]{\includegraphics[width=0.48\hsize]{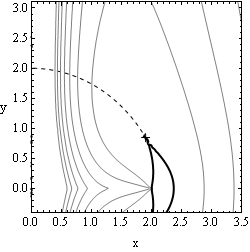}}
\caption{Evolution of the energy boundary function $E_{\rm b}(x,y;J,\omega,a)$ (black thick curve)  around the off-equatorial minima (denoted by black cross) demonstrated for fixed parameters $a=2$, $\omega=0$. Contours of the $E_{\rm b}(x,y;J,\omega,a)$ function are represented for various energy levels by gray curves; the ring singularity is located at the point $[a,0]$.
\label{offEQmin}
}
\end{figure}

%
\subsection{Off-equatorial extrema} \label{ExOffEqPlane}
%

For physical problems with spherical symmetry, e.g. the motion of test particles around \Schw{} BHs, it is meaningless to consider extrema of effective potential off the equatorial plane since the motion is always confined to a central plane. The physical situations related to string loops have axial  symmetry only, even if the background is spherically symmetric, so we have to consider also extrema  located off the equatorial plane. In fact, such off-equatorial extrema, minima only to be specific, appear in the braneworld spherically symmetric naked singularity spacetimes \citep{Stu-Kol:2012:JCAP:}.
To find $(r,\theta)$ position of the off-equatorial extrema in the Kerr spacetimes, we have to solve the relation 
\beq
\JJr^2(r,\theta;\omega,a) = \JJt^2(r,\theta;\omega,a) ; \label{offEX2}
\eeq
of course, we have to consider only positive values of $\JJr^2, \JJt^2$ functions determined by Eqs. (\ref{JExR}-\ref{JExT}). Unfortunately, the relation (\ref{offEX2}) implies a very complex function of $r$ and $\theta$ coordinates that gives only implicitly the solutions, assuming the parameters $\omega$ and $a$ fixed. Analytical formula for location of the off-equatorial extrema has not been found, the solutions are determined by numerical methods.

Due to the effect of gravitational attraction, the lowest part of the $E_{\rm} (x,y;J,\omega,a)$ function is located near the origin of coordinates. The effect of string current $J$ implies that the energy boundary function $E_{\rm b}$ has a "valley" located at $x \sim J$ for all values of coordinate $y$. 
The maximum of the $E_{\rm b}(x,y;J,\omega,a)$ function located at the ring singularity combined with the effect of the angular momentum $J$ barrier that always gives a valley at $x~\sim~J$, imply possibility of the off-equatorial minima. Numerical solutions (plotted in Fig. \ref{offEQex}) demonstrate that there are no off-equatorial extrema in the case of Kerr BHs ($0<a<1$), while in the case of Kerr NSs ($a>1$) only minima can exist. Due to the mirror symmetry of the energy boundary function $E_{\rm b}(x,y;J,\omega,a)$, the off-equatorial minima always appear in pairs - above and below the equatorial plane. As can be seen from Fig. \ref{offEQex}, the energy boundary function $E_{\rm b}(r,\theta;J,\omega,a)$ cannot have the off-equatorial minima for all values of $\theta$ coordinate; there exist some limiting value $\theta_{\rm end}(a)<\pi/2$ that can be determined numerically.

In Fig. \ref{offEQmin} we demonstrate dependence of the location of the off-equatorial minima on the latitudinal coordinate $\theta$ for the string loops with $\omega=0$ and give the related "lakes" of the constant energy sections of the energy boundary function $E=E_{\rm b}(x,y;J,\omega,a)$ governing trapped string loop states extending off the equatorial plane and their transition to states enabled to cross the equatorial plane or to escape to infinity. We have shown in Fig. \ref{offEQmin}(a) that for small values of the angular momentum $J$ and energy $E$ a closed string motion boundary around the off-equatorial minimum is formed close to the $y$-axis. If the energy $E$ is increased above $2J$ (Fig. \ref{offEQmin}(b)), the boundary becomes to be opened, allowing the string loop to cross the equatorial plane or escape to infinity. The off-equatorial "lake" pairs can exist for $\theta<\theta_{\rm end}$ (Fig. \ref{offEQmin}(c)), but for $\theta>\theta_{\rm end}$ they merge forming a "lake" crossing the equatorial plane (Fig. \ref{offEQmin}(d)).

\begin{figure*}
\subfigure[ \quad $a=0$]{\includegraphics[width=0.23\hsize]{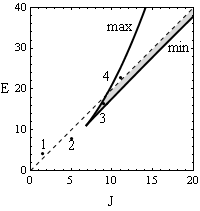}}
\subfigure[ \quad $a=0.99, \omega=-1$]{\includegraphics[width=0.23\hsize]{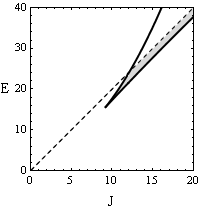}}
\subfigure[ \quad $a=0.99, \omega=0$]{\includegraphics[width=0.23\hsize]{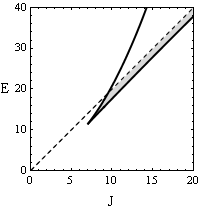}}
\subfigure[ \quad $a=0.99, \omega=+1$]{\includegraphics[width=0.23\hsize]{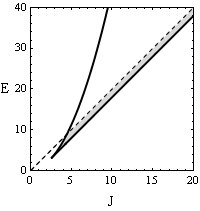}}
\caption{Extrema of the boundary energy function $E_{\rm b}(x,y;J,\omega,a)$ illustrated for \Schw{} and Kerr BH spacetimes as functions of the angular momentum $J$. Thick solid curves correspond to the maxima and minima of the energy boundary function $E_{\rm b}(x,y;J,\omega,a)$ in the equatorial plane. Minimum of the $E_{\rm b}$ function for the flat spacetime $E = 2J$ (escape to infinity) is represented by the dashed line. Regions where the ``lakes'' corresponding to the trapped states can exist are shaded. Numbers in Fig. (a) denote the examples of different types of the characteristic $E=$~const sections of the boundary energy function $E_{\rm b}(x,y;J,\omega,a)$, illustrated in Fig. \ref{MtypBH}.
\label{JEklasBH}
}
\end{figure*}

%
\section{Classification of the string loop dynamics in the Kerr BH and Kerr NS spacetimes}
%

\subsection{Black Holes}

There are four different types of the behavior of the energy boundary function for the string loop dynamics in the Kerr BH or \Schw{} spacetimes represented by the characteristic $E={\rm{}const}$ sections of the function $E_{\rm b}(r,\theta)$ in dependence on parameters $J,\omega,a$ \cite{Jac-Sot:2009:PHYSR4:}. We can distinguish them according to two properties: possibility of the string loop to escape to infinity in the $y$-direction, and possibility to collapse to the black hole. A detailed discussion can be found in \cite{Kol-Stu:2010:PHYSR4:}; here we shortly summarize the results. 

\begin{figure}
\includegraphics[width=0.5\hsize]{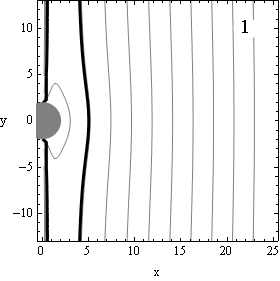}\includegraphics[width=0.5\hsize]{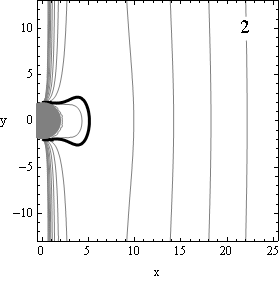}
\includegraphics[width=0.5\hsize]{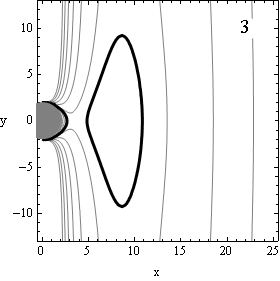}\includegraphics[width=0.5\hsize]{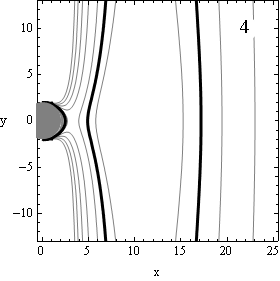}
\caption{
Constant energy sections of the energy boundary function $E=E_{\rm b}(x,y;J,\omega,a)$ governing the string loop dynamics in the Kerr or \Schw{} BH  spacetimes. Four different types of the behavior of the string loop dynamics are possible: collapse or escape to infinity (type 1), collapse to the BH (type 2), trapping in some "lake"-like region (type 3), escape to infinity without possibility to collapse to the BH (type 4), see Fig. \ref{JEklasBH}.
\label{MtypBH}
}
\end{figure}

The energy boundary function $E_{\rm b}(r,\theta;J,\omega,a)$ has two extrema, maximum and minimum, located above the black-hole horizon, if 
\beq
   J > J_{\rm E(min)}.
\eeq
For $J=J_{\rm E(min)}$, the energy boundary function $E_{\rm b}$ has an inflex point. For $J<J_{\rm E(min)}$ there are no extrema of the energy boundary function above the horizon. 

In the Kerr BH case, only the equatorial extrema occur. Using the formulae for the $E_{\rm b}(r;J,\omega,a)$ function and denoting the extrema by the relations 
\beq
  E_{\rm b(min)}= E_{\rm b}(r_{\rm E(min)}), \quad E_{\rm b(max)}= E_{\rm b}(r_{\rm E(max)}), 
\eeq
we give the extremal values of the boundary energy function in dependence on the string parameter $J>J_{\rm E(min)}$ in Fig. \ref{JEklasBH}.
The oscillatory motion in the $x$-direction is allowed for string loops with $J>J_{\rm E(min)}$ and energy satisfying the condition $E_{\rm b(min)}(J)<E<E_{\rm b(max)}(J)$. String loops with $J<J_{\rm E(min)}$, or with $E<E_{\rm b(max)(J)}$ and $J>J_{\rm E(min)}$, can be captured by the black hole.

\begin{figure*}
\subfigure[ \quad $a=1.1, \omega=-1$]{\includegraphics[width=0.26\hsize]{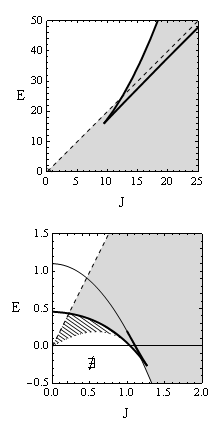}}
\hspace{-0.6cm}
\subfigure[ \quad $a=1.1, \omega=0$]{\includegraphics[width=0.26\hsize]{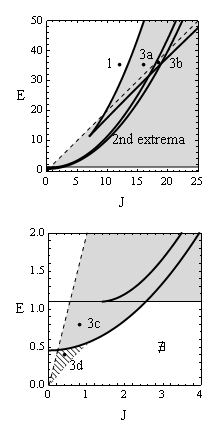}}
\hspace{-0.6cm}
\subfigure[ \quad $a=1.2, \omega=0$]{\includegraphics[width=0.26\hsize]{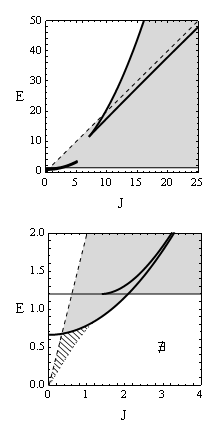}}
\hspace{-0.6cm}
\subfigure[ \quad $a=1.1, \omega=+1$]{\includegraphics[width=0.26\hsize]{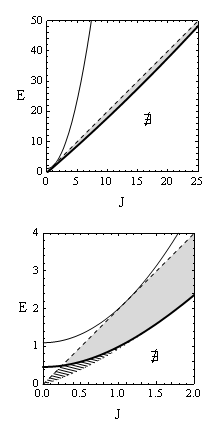}}
\caption{Extrema of the boundary energy function $E_{\rm b}(x,y;J,\omega,a)$ illustrated for the Kerr NS spacetimes as functions of the angular momentum $J$. The upper row of figures represents the large scale behavior, while the lower row figures, constructed for small values of $E$ and $J$, demonstrate details representing the off-equatorial minima. Thick solid curves correspond to the maxima and minima of the energy boundary function $E_{\rm b}$ in the equatorial plane. The boundary for the string loop escape to infinity along the $y$-axis, given by the relation $E=2J$, is illustrated by the dashed line.
Shaded are the regions where the constant energy sections of the energy boundary function $E=E_{\rm b}(x,y;J,\omega,a)$ form closed curves ("lakes") around minima of the function in the equatorial plane; hatched are the regions representing the off-equatorial "lakes" that exist around the off-equatorial minima of the energy boundary function. The off-equatorial minima can exist only for string loops with sufficiently small values of $J$ and $E$. Thin solid curve represents the limiting values of the energy boundary function just below the ring singularity, i.e. $E_{\rm b}(r\rightarrow~0,\pi/2;J,\omega,a)$; such values could correspond in the Kerr NS spacetimes to the lowest allowed values of the energy $E$ of the string loops with $\omega\in\{-1,0\}$. There is no maximum of the energy boundary function $E_{\rm b}$ in the case of $\omega=1$; negative energies are allowed only for $\omega=-1$. 
Regions of the forbidden combinations of the $E$ and $J$ parameters are represented by the symbol $\nexists$.  
The numbers correspond to the characteristic examples of the constant energy sections of the $E_{\rm b}(x,y;J;\omega,a)$ function as illustrated in Fig \ref{MtypNS}. 
\label{JEklasNS}
}
\end{figure*}

In the Kerr BH spacetimes, we can distinguish four different types of the behavior of the boundary energy function and the character of the string loop motion; in Figs. \ref{JEklasBH}, \ref{MtypBH} we denote them by points numbered by 1 to 4. 
The first case, $J~<~J_{\rm L2}$, corresponds to no inner and outer boundary - the string loop can be captured by the black hole or escape to infinity. The second case, $J_{\rm L2}~<~J~<~J_{\rm (e)eq}$, corresponds to the situation with an outer boundary - the string loop must be captured by the black hole. The third case, $J_{\rm (e)eq}~<~J~<~J_{\rm L1}$, corresponds to the situation when both inner and outer boundary exist - the string loop is trapped in some region forming a potential ``lake'' around the black hole. The fourth case, $J_{\rm L1}~<~J$, corresponds to an inner boundary - the string loop cannot fall into the black hole but it must escape to infinity. 

The main effect of the rotation parameter $a$ on the behavior of the energy boundary function $E_{\rm b}(x,y;J,\omega,a)$ in the Kerr BH spacetimes is shifting of the local extrema of the $E_{\rm b}$ function away ($\omega=-1$) or towards ($\omega=1$) the origin of coordinates, see Fig. \ref{IMPradii}. This effect can be observed also in dependence on the value of the $J_{\rm E(min)}(a)$ demonstrated in Fig. \ref{picJE2}. Using the characteristic cases of $E$--$J$ dependence represented in Fig. \ref{JEklasBH}, behavior of the energy boundary function $E_{\rm b}(x,y;J,\omega,a)$ in the Kerr BH spacetimes can be sorted in the following way reflecting that only quantitative differences occur due to the dependence on the string loop parameter $\omega$ and the spacetime spin parameter $a$. We assume $a=0.99$ giving the corresponding loci of the equilibrium string loop states:

\begin{enumerate}[(a)]
%
\item {\it \Schw{} BH} ($a=0$) The minimum of the $\JJ^2_{\rm (E)e}(r)$ function, $J_{\rm E(min)}$, corresponds to the marginally stable string loop radius $s_{\rm ms}\sim 4.30$, see Fig. \ref{picJE2}(a). Summary of the $E$--$J$ dependence can be inferred from Fig. \ref{JEklasBH}(a).
\item {\it Kerr BH} ($\omega=-1$) The local extrema of the $E_{\rm b}(x,y;J,\omega,a)$ function and the radius $s_{\rm ms-}\sim{}5.87$ are shifted away from the origin of coordinates, see Fig. \ref{JEklasBH}(b).
\item {\it Kerr BH} ($\omega=0$) The local extrema of the $E_{\rm b}(x,y;J,\omega,a)$ function and the radius $s_{\rm ms0}\sim{}4.26$ are shifted slightly towards the origin of coordinates, Fig. \ref{JEklasBH}(c).
\item {\it Kerr BH} ($\omega=1$) The local extrema of the $E_{\rm b}(x,y;J,\omega,a)$ function and the radius $s_{\rm ms+}\sim{}1.40$ are significantly shifted towards the origin of coordinates, see Fig. \ref{JEklasBH}(d).
\end{enumerate}

%
\subsection{Naked Singularities}

For Kerr NS we have found in the previous section more complex behavior of the energy boundary function $E_{\rm b}(x,y;J,\omega,a)$, since we have demonstrated existence of the off-equatorial minima and additional minima in the equatorial plane located close to the ring singularity. The off-equatorial minima of the $E_{\rm b}(x,y;J,\omega,a)$ function can exist only for small values of parameters $J$ and $E$ (see the case 3d in Figs. \ref{JEklasNS}(b),\ref{MtypNS}), while the equatorial minima located close to the ring singularity can exist even for $E>2J$ (see the cases 3a and 3b in Figs. \ref{JEklasNS}(b),\ref{MtypNS}).
Since in the Kerr NS spacetimes no event horizon exists, only collapse to the ring singularity can occur; string loops are usually trapped in closed regions of the $E={\rm~const}$ sections of the energy boundary function $E_{\rm b}(x,y;J,\omega,a)$, or escape to infinity along the $y$-axis. Characteristic examples of the energy boundary sections $E=E_{\rm b}(x,y;J,\omega,a)$ governing the string loop dynamics in the Kerr NS spacetimes are given in Fig. \ref{MtypNS}.

Using the characteristic cases of the $E$-$J$ dependence represented in Fig. \ref{JEklasNS},
behavior of the energy boundary function $E_{\rm b}(x,y)$ in the Kerr NS spacetimes can be sorted according to the parameters $a$ and $\omega$ in following way:

\begin{figure}
\includegraphics[width=0.5\hsize]{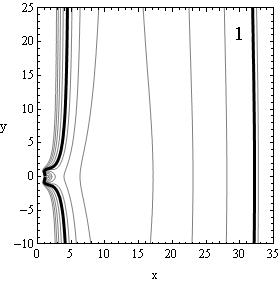}\includegraphics[width=0.5\hsize]{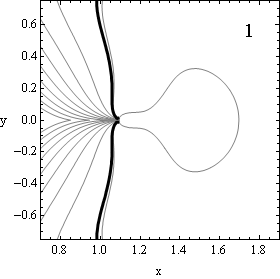}
\includegraphics[width=0.5\hsize]{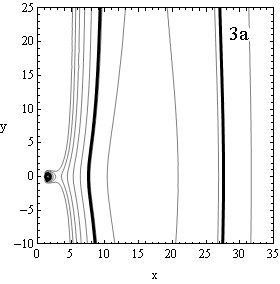}\includegraphics[width=0.5\hsize]{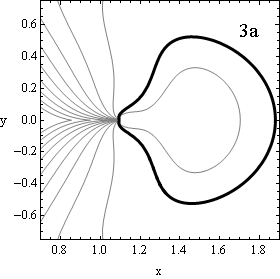}
\includegraphics[width=0.5\hsize]{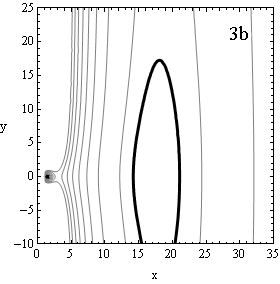}\includegraphics[width=0.5\hsize]{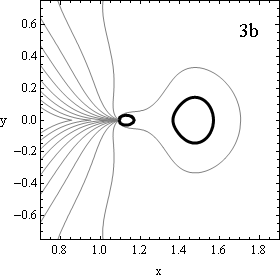}
\includegraphics[width=0.5\hsize]{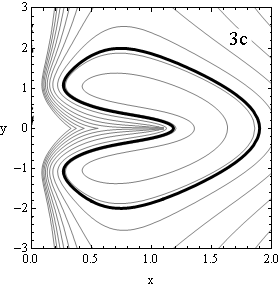}\includegraphics[width=0.5\hsize]{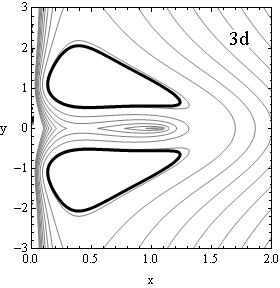}
\caption{Constant energy sections of the energy boundary function governing the string loop dynamics in the Kerr NS spacetimes. The string loops escape to infinity (type 1 and type 3a), or are trapped in closed boundaries ("lakes") formed around minima of the $E_{\rm b}(x,y;J,\omega,a)$ function in the equatorial plane (types 1 and 3a-3c), or are trapped out of the equatorial plane (3d) - see Fig. \ref{JEklasNS} for detailed classification. The "lakes" are approaching the ring singularity in the cases 1, 3a, and partially in the 3b case, while "lakes" completely separated from the ring singularity occur in the cases 3b, 3c and 3d. 
\label{MtypNS}
}
\end{figure}
%
\begin{enumerate}[(a)]
\item {\it Kerr NS} ($\omega=-1$) Function $\JJ^2_{\rm (E)e}(r)$, determining extrema of the $E_{\rm b}(r)$ function in the equatorial plane, has now two extrema, $J_{\rm E(min)}$ and $J_{\rm E(max)}$, see Fig. \ref{picJE2}(e). For $J<a$, the function $E_{\rm b}(r)$ has only a local minimum, corresponding to the lowest value of the string loop energy $E$. For $J<J_{\rm E (max)}$, the function $E_{\rm b}(r)$ has a local maximum and minimum in the vicinity of the ring singularity. For $J_{\rm E(max)}<J<J_{\rm E(min)}$, the function $E_{\rm b}$ has no local extrema, the string loop can obtain lowest possible energy $E$ near the ring singularity. The lowest string loop energy is given by Eq. (\ref{lowE}), and can take also negative values, see Fig. \ref{sectionEb}. For $J>J_{\rm E(min)}$, there exist the "standard" extrema of the $E_{\rm b}$ function, observed also in the case of Kerr BH, i.e., a maximum and minimum located close to $x\sim~J$. Summary of $E$--$J$ dependence is illustrated in Fig. \ref{JEklasNS}(a).
\item {\it Kerr NS} ($\omega=0, a<a_{\rm crit}$) The local extrema of the $E_{\rm b}(x,y;J,\omega,a)$ function are determined by the behavior of the $J^2_{\rm (E)eq}$ function, which has now only a minimum $J_{\rm E(min)}$, see Fig. \ref{picJE2}(f). Because for $r \rightarrow 0$, there is $J^2_{\rm (E)eq} = 2$, the function $E_{\rm b}$ has only one minimum located in vicinity of the ring singularity for $J<\sqrt{2}$, while for $J>\sqrt{2}$, an additional maximum of the $E_{\rm b}$ function occurs in the  vicinity of the ring singularity. For $J>J_{\rm E(min)}$, the standard extrema of the $E_{\rm b}$ function exist - a maximum and minimum, located close to $x\sim~J$. In this particular case, three separated closed areas governing the string loop dynamics occur in the $E=E_{\rm b}(x,y)$ sections - two of them occur near the ring singularity, and the third one close to $x\sim~J$, see the case 3b in Fig. \ref{MtypNS}.
\item {\it Kerr NS} ($\omega=0, a>a_{\rm crit}$) The only differences with respect to the previous case (b) arises in the behavior of the $J^2_{\rm (E)eq}$ function. Here, a new maximum $J_{\rm E(max)}$ occurs,  hence, some extrema of the $E_{\rm b}$ function do not exist for all values of the parameter $J$. Minimum of the $E_{\rm b}$ function (and the second "lake") located close to the ring singularity  disappear for $J>J_{\rm E(max)}$, while this second "lake" and the "lake" located close to $x=J$ can exist simultaneously only if $J_{\rm E(max)} > J_{\rm E(min)}$, see Fig. \ref{JEklasNS}(c).
\item {\it Kerr NS} ($\omega = 1$) Local extrema of the $E_{\rm b}$ function in the equatorial plane are determined by the $J^2_{\rm(E)eq}$ function which has no extrema in this case - for any value of the parameter $J$, the energy boundary function $E_{\rm b}$ has only one minimum (and the  corresponding "lake"), see Fig. \ref{picJE2}(h). Of course, as typical for Kerr NS, there exist the off-equatorial minima and related "lakes" of the constant energy sections for small angular momenta $J$ and energies $E$, see Fig. \ref{JEklasNS}(d).
\end{enumerate}

%
\section{String loop acceleration and asymptotical ejection speed}
%

From the astrophysical point of view, one of the most relevant applications of the axisymmetric string loop motion is the possibility of strong acceleration of the linear translational string loop motion due to the transmutation process in the strong gravity of extremely compact objects that could well mimic acceleration of relativistic jets in Active Galactic Nuclei (AGN) and microquasars \cite{Jac-Sot:2009:PHYSR4:,Stu-Kol:2012:PHYSR4:,Stu-Kol:2012:JCAP:}. Since the Kerr BH and NS spacetimes are asymptotically flat, we first discuss the string loop motion in the flat spacetime that enables clear definition of the acceleration process. The energy of the string loop (\ref{StringEnergy}) in the flat spacetime,  expressed in the Cartesian coordinates, reads
\beq
E^2 = \dot{y}^2 + \dot{x}^2 + \left( \frac{J^2}{x} + x \right)^2  =  E^2_{\mathrm y} + E^2_{\mathrm x}, \label{E2flat}
\eeq
where dot denotes derivative with respect to the affine parameter $\af$. The energy related to the motion in the $x$- and $y$-directions are given by the relations
\beq
  E^2_{\mathrm y} = \dot{y}^2, \quad E^2_{\mathrm x} = \dot{x}^2 + \left( \frac{J^2}{x} + x \right)^2 = (x_{\rm i} + x_{\rm o})^2 = E^2_{0}  \label{restenergy}
\eeq
where $x_i$ ($x_o$) represent the inner (outer) limit of the oscillatory motion. The energy $E_0$ representing the internal energy of the string loop is minimal when the inner and the outer radii coincide, leading to the relation 
\beq
 E_{\rm 0(min)} = 2 J \label{E0min}
\eeq 
that determines the minimal energy necessary for escaping of the string loop to infinity in the Kerr BH or NS spacetimes. Clearly, $E_{\rm x}=E_{0}$ and $E_{\rm y}$ are constants of the string loop motion and no transformation between these energy modes is possible in the flat spacetime. However, in strong gravity in vicinity of black holes or naked singularities, the internal kinetic energy of the oscillating string can be transmitted into the kinetic energy of the translational linear motion (or vice versa) due to the chaotic character of the string loop dynamics \cite{Jac-Sot:2009:PHYSR4:,Stu-Kol:2012:PHYSR4:}. 

\begin{figure}
\includegraphics[height=0.64\hsize]{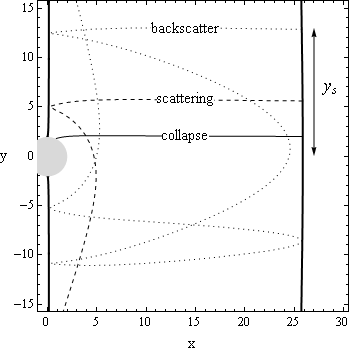}
\includegraphics[height=0.65\hsize]{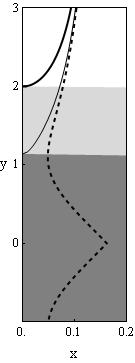}
\caption{Comparison of the energy boundary function $E_{\rm b}$ for characteristic values of the spin parameter $a=0$ (thick solid), $a=0.99$ (black solid), $a=1.1$ (dashed curve). The functions $E_b$  differ significantly near the origin of the coordinates for given cases; we also depict area of the  horizon for black holes (gray). The string loop motion is represented for appropriately chosen characteristic cases. Because of the sensitivity of the string loop motion to the initial conditions, we have to compare whole set of trajectories to see the effect of the rotation parameter $a$. We "shoot" the string loops from some position $x_s \sim 25, y_s \in (0,13)$, with current $J=2$ and energy $E=25$ towards the BH (NS). We continuously change the coordinate $y_s$ which serves as an impact parameter and plot the resulting gamma factor; results for different $a$ and $\omega$ are given in Fig. \ref{stringFIG_5}.
\label{schemaBHvsNS} \label{schemaACC}}
\end{figure}

\begin{figure*}
\includegraphics[width=0.95\hsize]{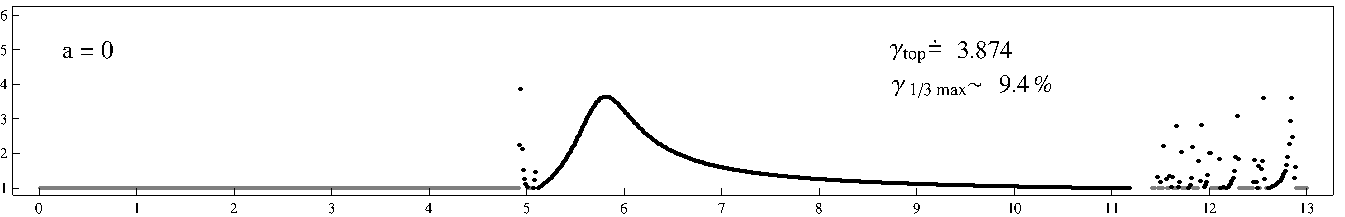}
\includegraphics[width=0.95\hsize]{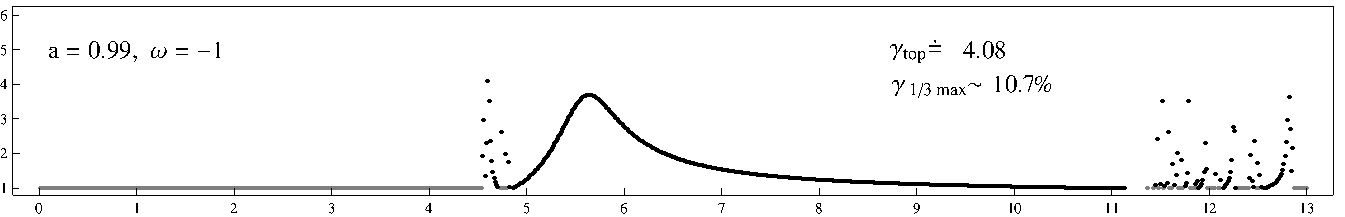}
\includegraphics[width=0.95\hsize]{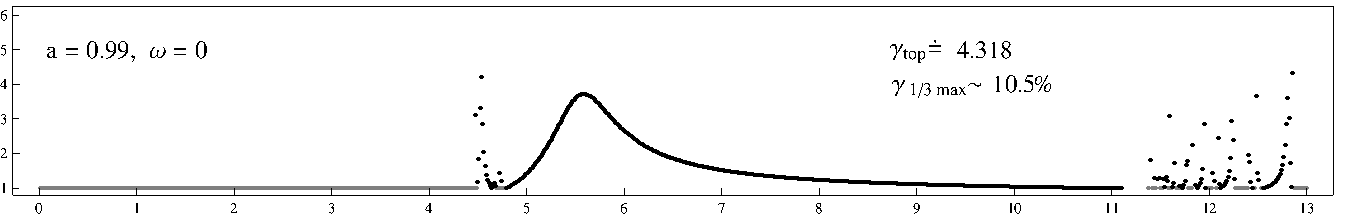}
\includegraphics[width=0.95\hsize]{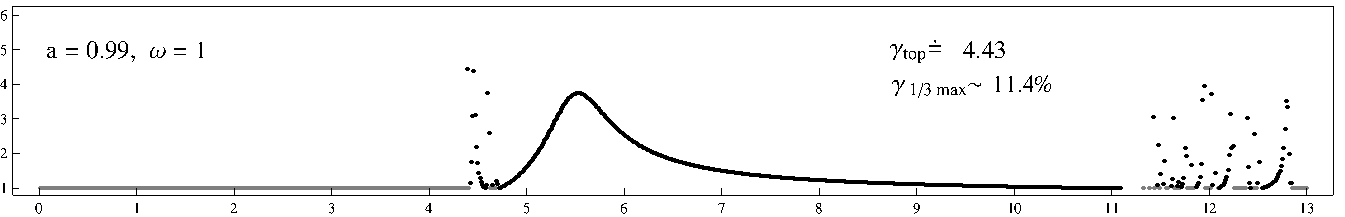}
\includegraphics[width=0.95\hsize]{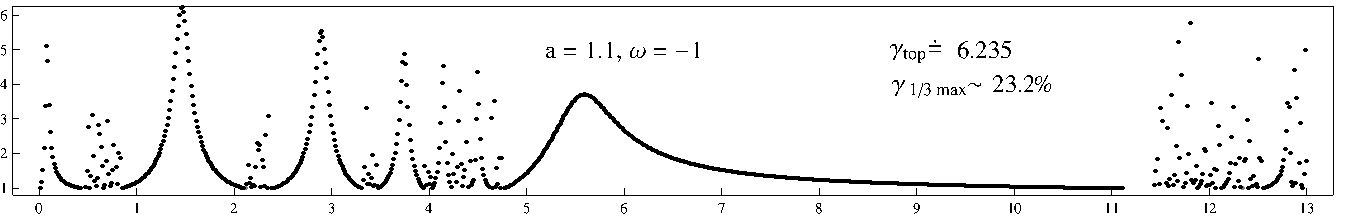}
\includegraphics[width=0.95\hsize]{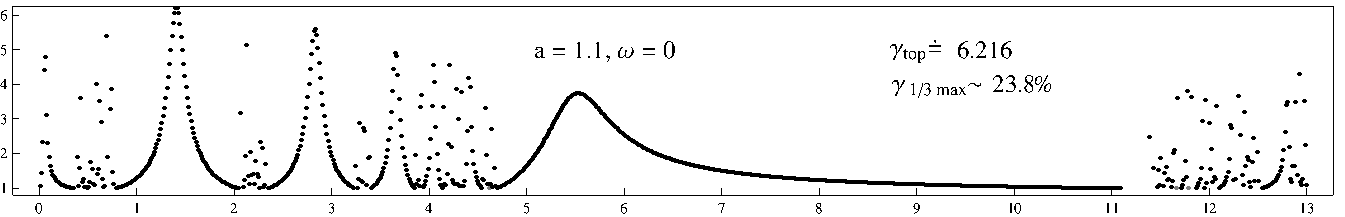}
\includegraphics[width=0.95\hsize]{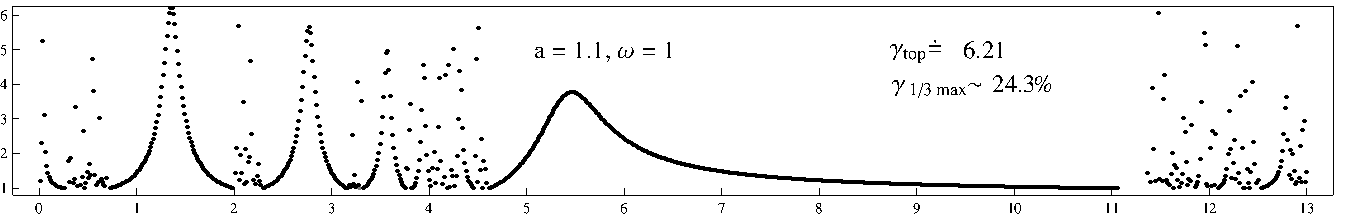}
\vspace{-0.3cm}
\caption{\label{stringFIG_5}
The asymptotic speed of transmitted string loops given for all three values of the string loop parameter $\omega$ and characteristic values of the spacetime spin parameter $a$. The Lorentz factor $\gamma$ (vertical axis) is calculated for string loops with energy $E=25$ and current $\JJ=2$, starting from the rest with different initial position $\yy_0 \in (0,13) $ (horizontal axis) while $\xx_0$ is calculated from (\ref{StringEnergy}). Gray points correspond to the string loops collapsed to the black hole. Maximal acceleration (\ref{gmmax}) for this case gives us the limiting gamma factor $\gamma_{\rm max} = 6.25$. We show the topical gamma factor that is numerically found in the sample, $\gamma_{\rm top}$, and also the efficiency of the transmutation effect, $R_{\gamma(\rm max)/3}$, given by the relative number of accelerations when the final Lorentz factor is larger than $\gamma_{\rm max}/3$.
}
\end{figure*}

\begin{figure}
\subfigure[\quad $a=0$]{\includegraphics[width=0.48\hsize]{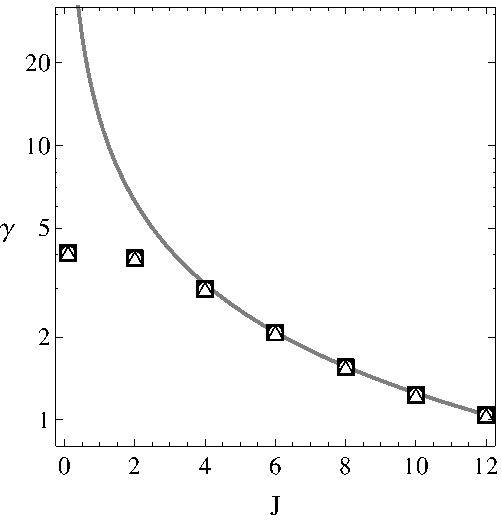}}
\subfigure[\quad $a=0.99$]{\includegraphics[width=0.48\hsize]{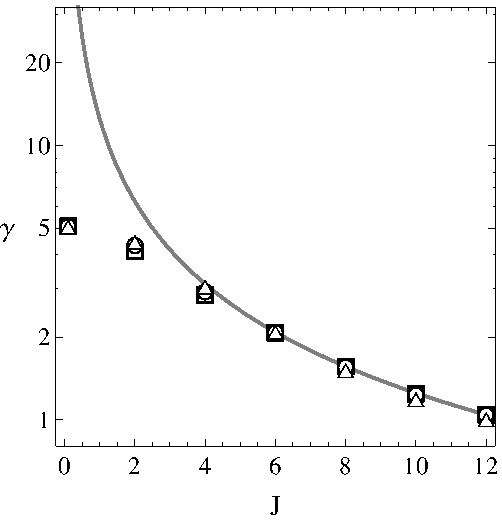}}
\subfigure[\quad $a=1.1$]{\includegraphics[width=0.48\hsize]{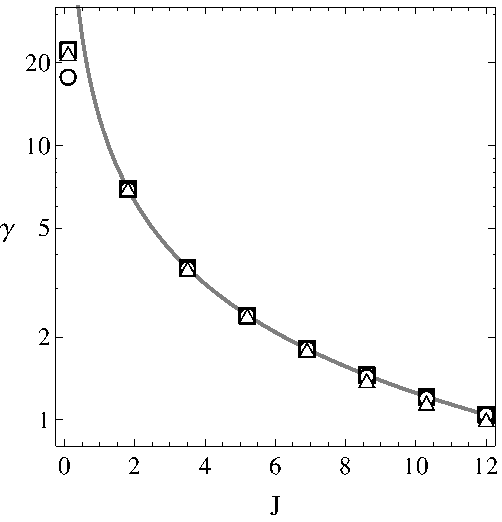}}
\subfigure[\quad $a=2$]{\includegraphics[width=0.48\hsize]{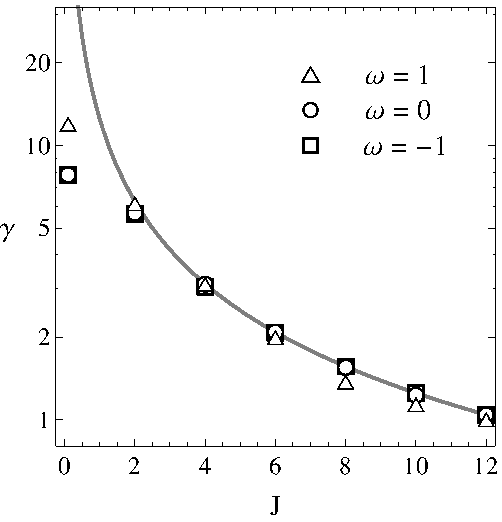}}
\caption{
Maximal and extremal (topical) Lorentz gamma factor $\gamma$ in dependence on the angular momentum parameter $J$ for fixed string loop energy $E=25$, parameter $\omega=-1,0,+1$, and the spin parameter $a$. Solid curve shows the predicted maximal acceleration (given by the theoretical limit) represented by the Lorentz factor $\gamma_{\rm max}$ in dependence on $J$, eq. (\ref{gmmax}), while the markers are numerically calculated extremal (topical) Lorentz factors $\gamma_{\rm top}$ from the sample under consideration.
}
\end{figure}

In order to get a strong acceleration in the Kerr BH and NS spacetimes, the string loop has to pass the region of strong gravity near the black hole horizon or in vicinity of the naked ring singularity, where the string transmutation effect $E_{\rm x} \leftrightarrow E_{\rm y}$ can occur. All energy of the transitional ($E_{\rm y}$) energy mode can be transmitted to the oscillatory ($E_{\rm x}$) energy mode - oscillations of the string loop in the $x$-direction and the internal energy of the string will increase maximally in such a situation, while the string will stop moving in the $y$-direction. However, all energy of the $E_{\rm x}$ mode cannot be transmitted into the $E_y$ energy mode - there remains inconvertible internal energy of the string,  $E_{\rm 0(min)}=2J$, being the minimal potential energy hidden in the $E_{\rm x}$ energy mode. 

The final Lorentz factor of the transitional motion of an accelerated string loop as observed in the asymptotically flat region of the Kerr spacetimes is, due to (\ref{restenergy}), determined by the relation \citep{Jac-Sot:2009:PHYSR4:,Stu-Kol:2012:PHYSR4:} 
\beq
 \gamma = \frac{E}{E_0} = \frac{E}{\xx_{\rm i} + \xx_{\rm o}}, \label{gamma}
\eeq 
where $E$ is the total energy of the string loop moving with the internal energy $E_{0}$ in the $\yy$-direction with the velocity corresponding to the Lorentz factor $\gamma$. Clearly the maximal Lorentz factor of the transitional motion then reads \cite{Stu-Kol:2012:PHYSR4:}
\beq
 \gamma_{\rm max} = \frac{E}{2 J} \label{gmmax}.
\eeq
From this equation we see that large ratio of the string loop energy $E$ versus its angular momentum  $J$ is needed for ultra-relativistic acceleration. It should be stressed that rotation of the black hole (naked singularity) is not a relevant ingredient of the acceleration of the string loop motion due to the transmutation effect \cite{Stu-Kol:2012:PHYSR4:}, contrary to the Blandford---Znajek effect \cite{Bla-Zna:1977:MNRAS:} usually considered in modeling acceleration of jet-like motion in AGN and microquasars.

\begin{figure}
\includegraphics[width=\hsize]{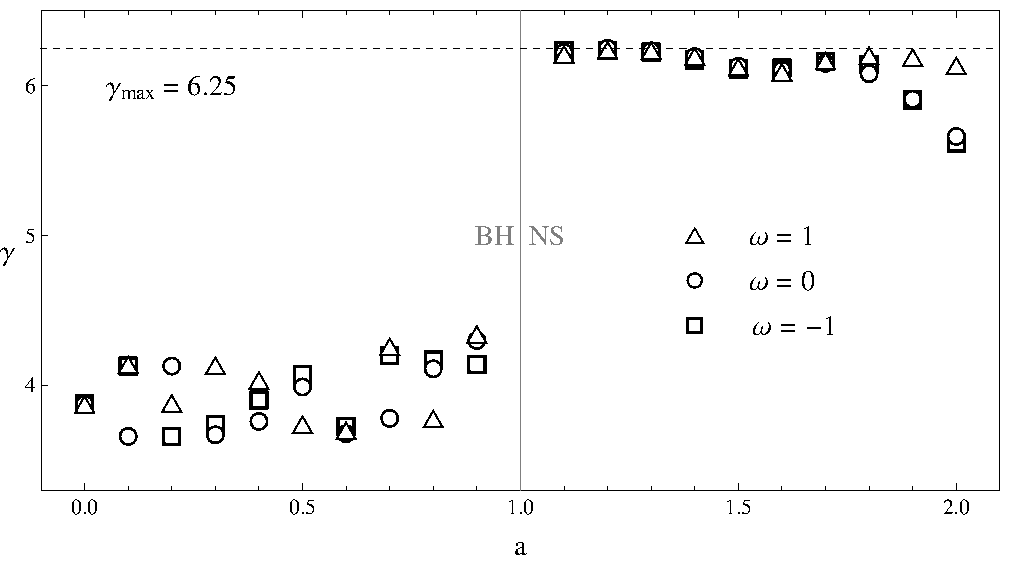}
\caption{\label{stringFIG_9}
Topical Lorentz factors $\gamma_{\rm top}$ given for all three possible values of $\omega = -1, 0, +1$ of string loops in dependence on the spin parameter $a$. Notice the sharp change at the BH/NS transition. String loops are assumed with energy $E=25$ and angular momentum $\JJ=2$, maximal acceleration is limited by $\gamma_{\rm max} = 6.25$.}
\end{figure}

\begin{figure*}
\subfigure[\quad $J=0.1$]{\includegraphics[width=0.49\hsize]{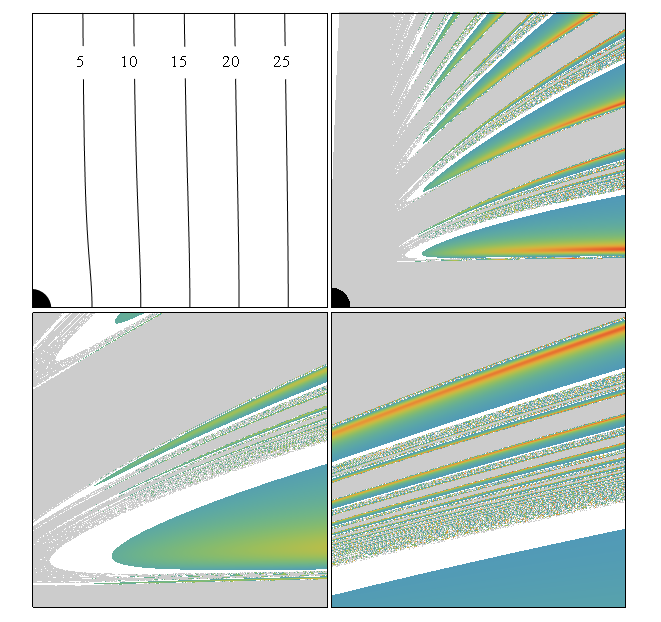}}
\subfigure[\quad $J=11$]{\includegraphics[width=0.49\hsize]{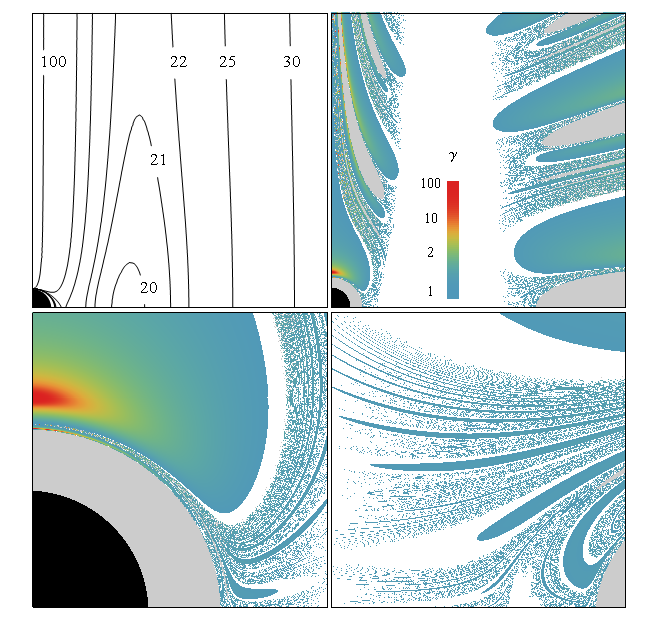}}
\caption{
String loop acceleration in the \Schw{} BH spacetime, plotted for various starting points of the string loop and its energy. The string loop is starting from the rest at the points $x_{\rm s} \in (0.1, 30.1), y_{\rm s} \in (0.1, 30.1)$ with the angular momentum fixed to values of $J=0.1$ or $J=11$, but with different energy $E$ that is determined by the starting point. The energy can be estimated from the first subfigure, where different levels of the $E_{\rm b}(x,y;J,\omega,a)$ function are plotted. 
The presented figures correspond to the Fig. 22 from \cite{Kol-Stu:2010:PHYSR4:}, but here we colored every point according to the asymptotic Lorentz factor of the translational string loop motion in the $y$-direction. Black color denotes region below the horizon; gray regions correspond to the string loops collapsed to the black hole. Regions of white color correspond to string loops that do not reach "infinity" located at $r=1000$ for given time $\af=200$ and remain oscillating around the BH.
\label{niceFIG1}
}
\vspace{0.5cm}
\begin{minipage}[b]{\hsize}
\subfigure[\quad $J=0.1, \omega=0$]{\includegraphics[width=0.255\hsize]{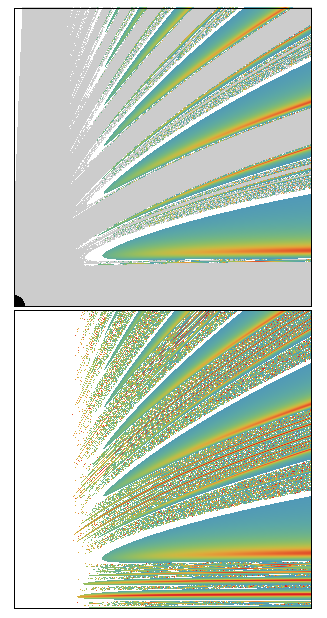}}
\hspace{-0.017\hsize}
\subfigure[\quad $J=11, \omega=-1$]{\includegraphics[width=0.255\hsize]{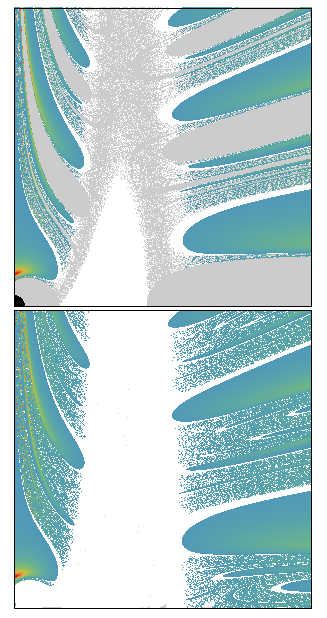}}
\hspace{-0.03\hsize}
\subfigure[\quad $J=11, \omega=0$]{\includegraphics[width=0.255\hsize]{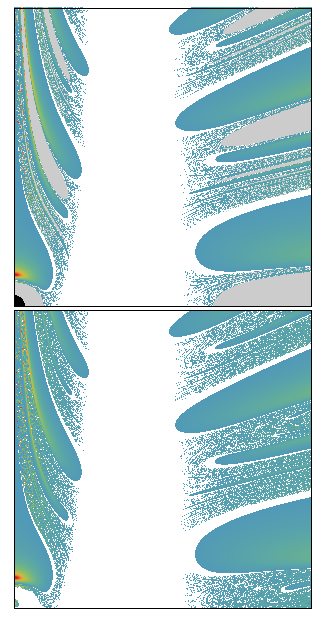}}
\hspace{-0.03\hsize}
\subfigure[\quad $J=11, \omega=1$]{\includegraphics[width=0.255\hsize]{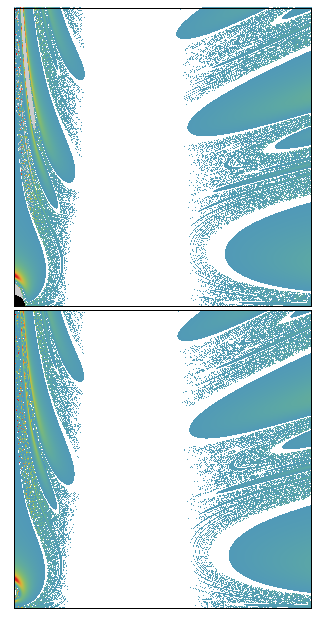}}
\end{minipage}
\caption{
String loop Lorentz factors of accelerated motion in the Kerr BH ($a=0.99$), upper row, and Kerr NS ($a=1.1$), lower row, spacetimes for $J=0.1$ and $J=11$ angular momenta; different values of the parameter $\omega$ are considered. The starting positions of the string loop agree with those in Fig. \ref{niceFIG1} plotted for $a=0$, see description there. In the Kerr BH case with string loop parameter $\omega=-1$, we observe significantly extended region of the gray colour, leading to conclusion that the string loop configurations with $L<0$, ale less stable and tend to collapse to the Kerr BH more rapidly. In the Kerr NS spacetimes, the region of collapsed trajectories (gray) is substituted by the region of accelerated trajectories.
\label{niceFIG2}
}
\end{figure*}

For energy and angular momentum parameters of a string loop fixed, increasing rotation parameter $a$ causes only slight increase of the energy efficiency of the transmutation process (maximal Lorentz factor) in the Kerr BH in comparison to the Schwarzschild case, but we observe a substantial jump of the energy efficiency after crossing the region of Kerr NS, see Fig. \ref{stringFIG_9}. In Fig. \ref{stringFIG_5} we see that the string loops are accelerated in naked singularity gravitational fields almost to the  maximal possible acceleration $\gamma_{\rm max}=6.25$ for the energy $E=25$ and angular momentum $J=2$.
The physical reason for distinction of the transmutations process efficiency in the black hole and naked singularity spacetimes comes from the non-existence of an event horizon in the naked singularity spacetimes, see \cite{Stu-Kol:2012:JCAP:}. The transmutation efficiency is highest at the deepest parts of the gravitational field of black holes or naked singularities. After bouncing at the deepest part of the potential well of a black hole, the string loop crosses the event horizon and is captured by the black hole. On the other hand, string loops are not captured in the field of naked singularities, and can be accelerated to highest velocities. For naked singularities we see in Fig. \ref{stringFIG_5} that the most accelerated string loops with $\gamma_{\rm top} \doteq 6.2$ are starting near the equatorial plane ($y_s \in (0,5)$), however, such string loops are captured in the black hole cases (see Fig. \ref{stringFIG_5}). The situation of the transitions between the oscillatory motions and the transitional accelerated motion can be properly represented by their distribution in the space of initial states $x_{s} - y_{s}$  - see Figs \ref{niceFIG1} and \ref{niceFIG2}

The acceleration of the string loops works equally for both cosmic strings that could be remnants of some phase transitions in the early universe \cite{Wit:1985:NuclPhysB:,Vil-She:1994:CSTD:}, and for plasma demonstrating a string-like behavior either due to magnetic field line tubes captured in the plasma \cite{Sem-Dya-Pun:2004:Sci:}, or due to thin flux tubes of plasma that can be effectively described as a one-dimensional string \citep{Spr:1981:AA:,Cre-Stu:2013:PhRvE:}. We can expect that in such situations the relevant physics can be described by the string dynamics instead of much more complex magnetohydrodynamics governing plasma in general situations. 

The magnetized plasma string-like structures could be relevant in acceleration of collimated jets observed in accreting astrophysical systems ranging from young stars, stellar mass black holes or neutron stars, to supermassive black holes (or, alternatively, Kerr superspinars) in quasars and active galactic nuclei. As discussed in the pioneering work of Jacobson and Sotiriou \cite{Jac-Sot:2009:PHYSR4:}, processes with "magnetic" string loops could arise near the equatorial plane of  accreting systems and due to the transmutation process of converting the internal string energy to the kinetic energy of their translational motion a stream of string loops moving along the axis could appear representing thus a well collimated jet.

%
\section{Regular and chaotic motion}
%

\begin{figure*}
\subfigure[$\,\, E = 5 \,\,(\JJ \sim 3.5)$]{\includegraphics[width=0.23\hsize]{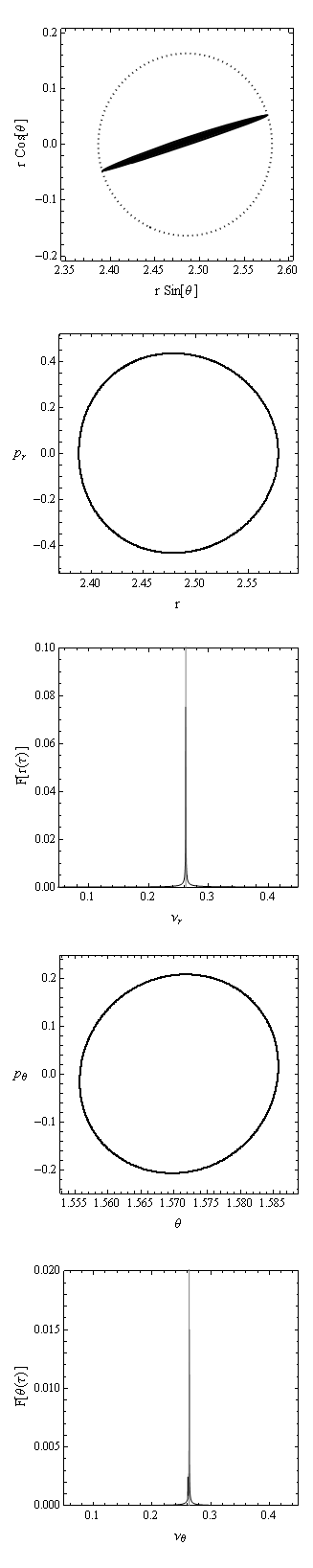}}
\subfigure[$\,\, E = 8 \,\,(\JJ \sim 4.7)$]{\includegraphics[width=0.23\hsize]{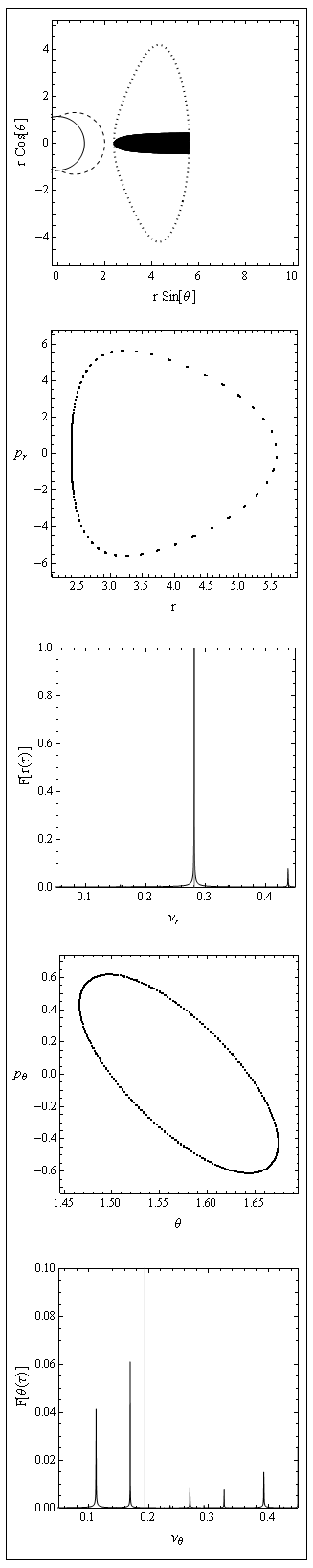}}
\subfigure[$\,\, E = 8.3 \,\,(\JJ \sim 4.8)$]{\includegraphics[width=0.23\hsize]{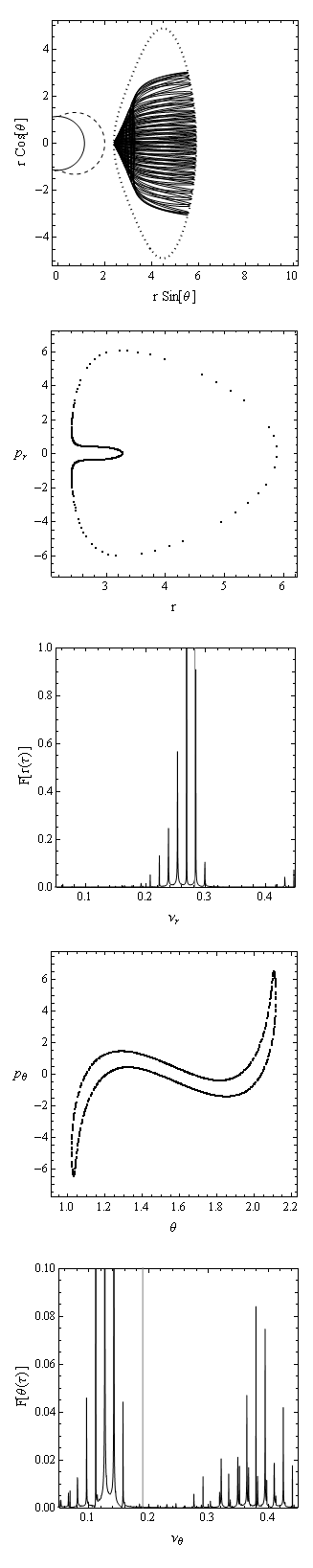}}
\subfigure[$\,\, E = 9   \,\,(\JJ \sim 5.1)$]{\includegraphics[width=0.23\hsize]{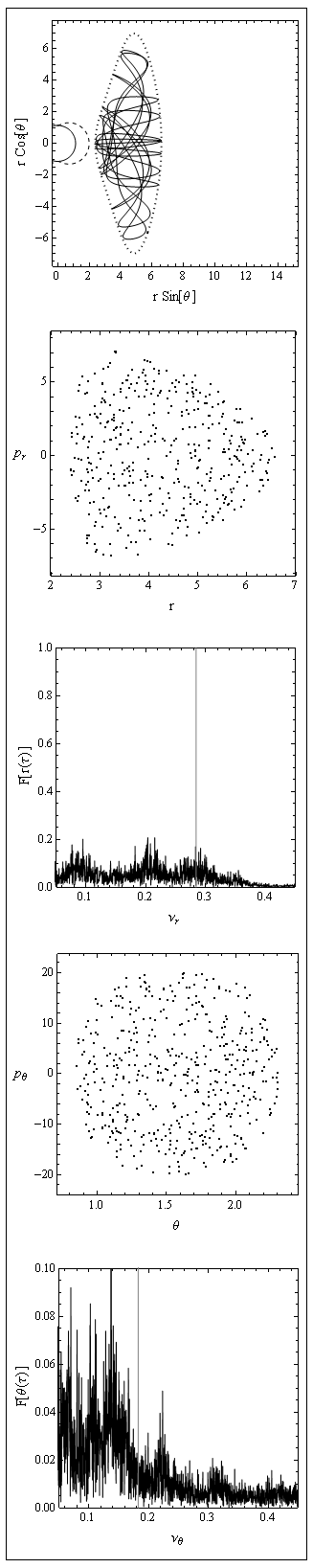}}
\caption{
Transition from regular to chaotic regime of the string loop motion as a solution to the "focusing" of trajectories problem \cite{Jac-Sot:2009:PHYSR4:}.
All four calculated cases (a-d) demonstrate increase of chaoticity as the string-loop energy $E$ grows. We use the Poincare sections of the $r-p_r$ ($\theta-p_\theta$) spaces, and the Fourier spectrum of the coordinates $r(\af)$ ($\theta(\af)$). The string loop is starting from the point $P_1$ ($r_0=r_{\rm h}+1.25,\theta_0=\pi/2 + 0.02$) in the Kerr BH with $a=0.99$ with various values of energy $E$ (and current $\JJ$). The parameter $\omega$ is fixed to be $\omega=1$. The vertical lines in the Fourier spectra are the frequencies $ \omega_r / (2 \pi), \omega_\tt / (2 \pi)$.
Boxed column of figures "regular" (b) and "chaotic" (d) are corresponding to the cases 9(a),8(a) from \cite{Jac-Sot:2009:PHYSR4:}.
\label{FIG_chaP1}
} 
\end{figure*}

\begin{figure*}
\subfigure[$\,\, E = 3.3 \,\,(\JJ \sim 2.7)$]{\includegraphics[width=0.23\hsize]{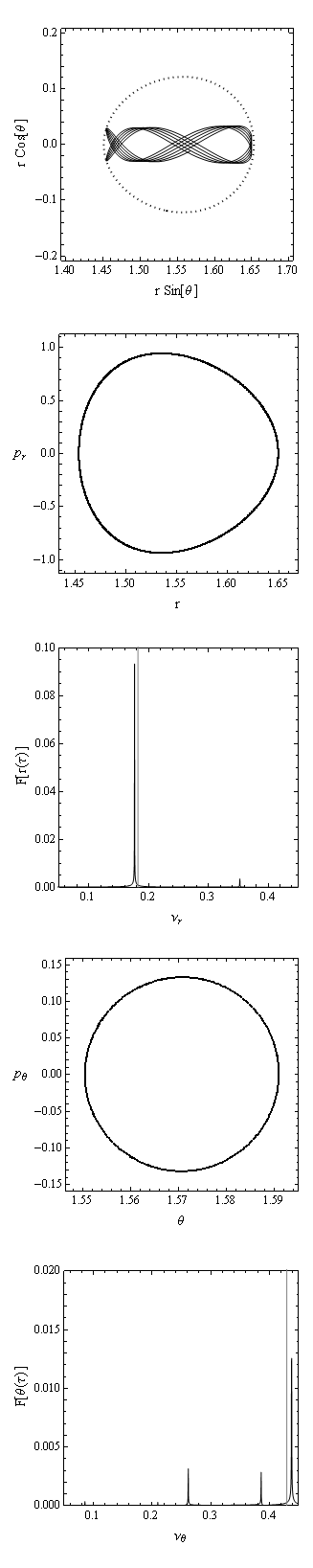}}
\subfigure[$\,\, E = 5.5 \,\,(\JJ \sim 3.5)$]{\includegraphics[width=0.23\hsize]{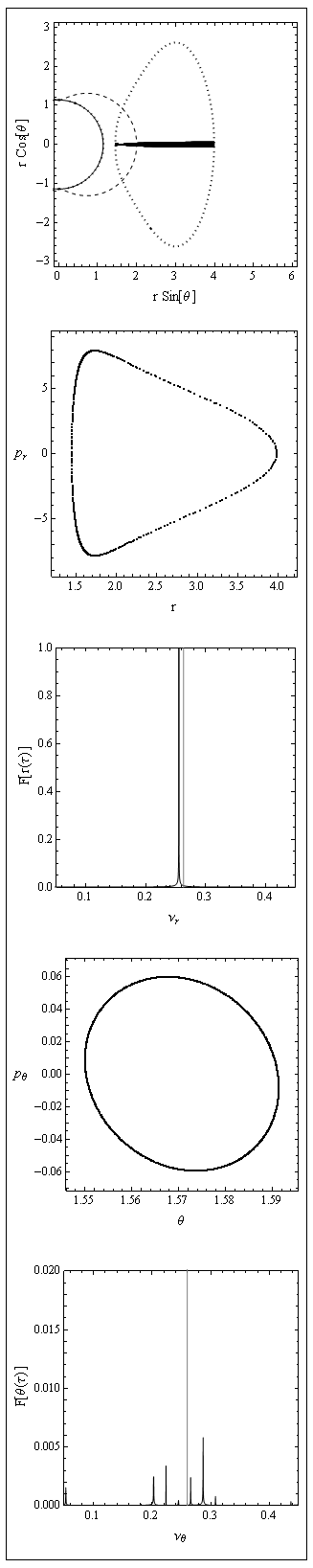}}
\subfigure[$\,\, E = 5.7 \,\,(\JJ \sim 3.6)$]{\includegraphics[width=0.23\hsize]{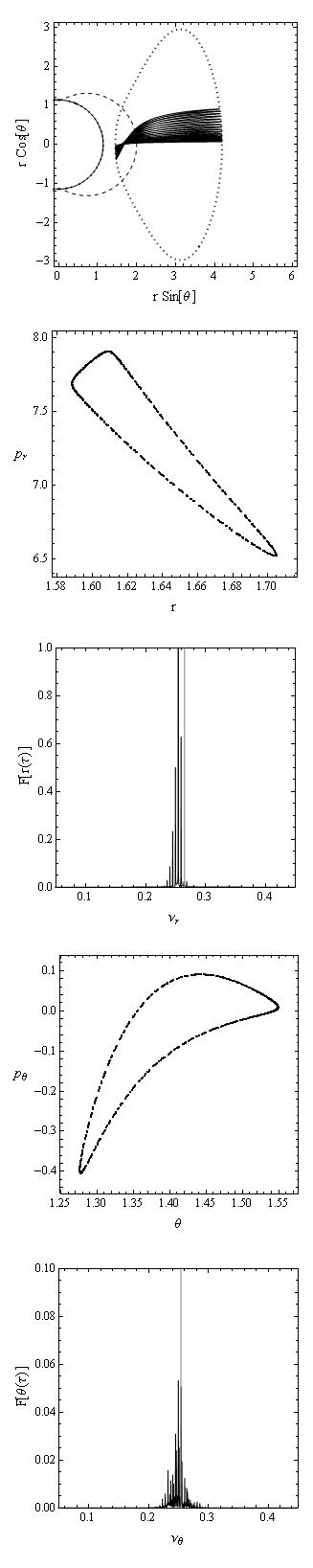}}
\subfigure[$\,\, E = 8   \,\,(\JJ \sim 4.3)$]{\includegraphics[width=0.23\hsize]{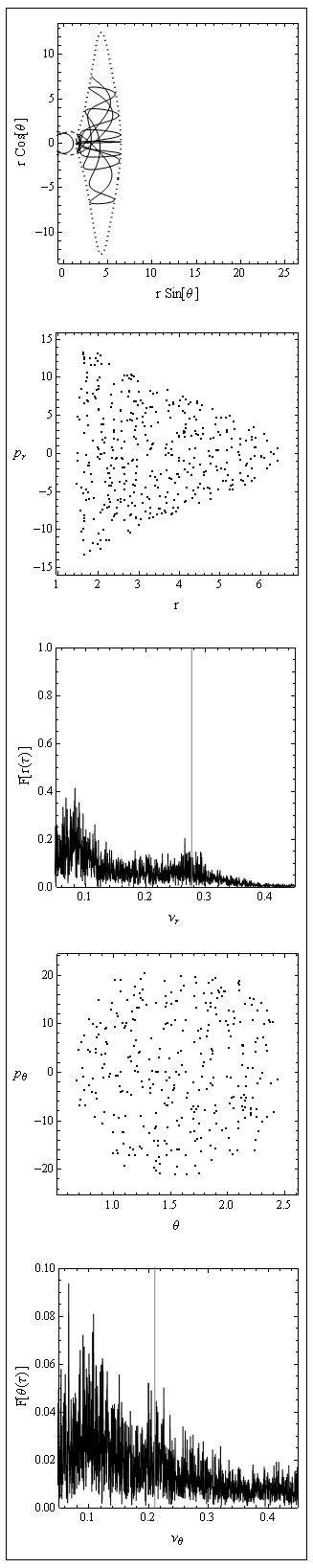}}
\caption{ 
Transition from regular to chaotic regime of the string loop motion. We use Poincare sections of the $r-p_r$ ($\theta-p_\theta$)space and the Fourier spectrum of the coordinates $r(\af)$ ($\theta(\af)$). The string loop is starting from the point $P_2$ ($r_0=r_{\rm ISCO},\theta_0=\pi/2 + 0.02$) in the Kerr BH with $a=0.99$ with various values of energy $E$ (and current $\JJ$). The parameter $\omega$ is fixed to be $\omega=1$.
Boxed column of figures "regular" (b) and "chaotic" (d) correspond to the cases 9(c),8(c) from \cite{Jac-Sot:2009:PHYSR4:}. There is no "focusing" of string trajectories in this case, but only increase of chaoticity of the motion.
\label{FIG_chaP2}
} 
\end{figure*}

Motion of strings is generally a chaotic motion, even in the case of axisymmetric string loops  \cite{Lar:1994:CLAQG:,Stu-Kol:2012:JCAP:}. Nevertheless, in the motion of the string loops can occur, in analogy with the motion of charged test particles \citep{Kop-Kar-Kov-Stu:2010:ASTRJ:,Kov-Kop-Kar-Koj:2013:CLAQG:,Sem-Suk:2012:MNRAS:}, "islands of regularity". We shall discuss the connection of the regular and chaotic motion of string loops in the Kerr BH and NS spacetimes around the stable equilibrium points of the motion. We can demonstrate that transitions from the regular to chaotic motion is of the same character in both black hole and naked singularity spacetimes.

\begin{figure*}
\subfigure[\quad point $P_1$]{\includegraphics[width=0.45\hsize]{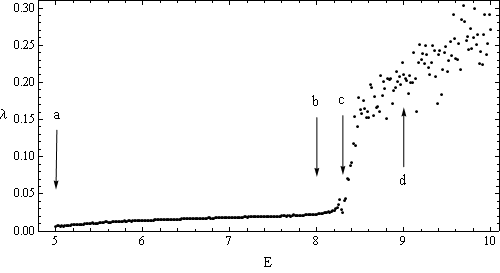}}
\hspace{0.5cm}
\subfigure[\quad point $P_2$]{\includegraphics[width=0.45\hsize]{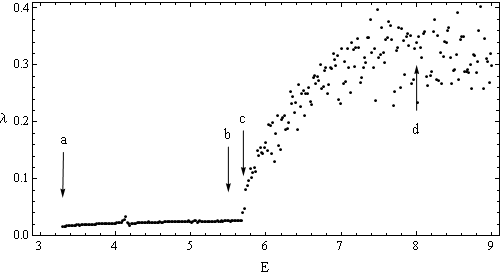}}
\caption{ 
Growing of the maximal Lyapunov exponent in dependence on the increasing string loop energy $E$. For small energies the motion is regular, but above some critical energy $E_{\rm c(P1)}\sim 8.2$ ($E_{\rm c(P2)}\sim 5.7$) the chaotic regime starts. Arrows with letters mark individual cases taken from Figs.\ref{FIG_chaP1},\ref{FIG_chaP2}.
\label{FIG_Lyap}
} 
\end{figure*}


The equilibrium points of the Hamiltonian (\ref{HamHam}) correspond to the local minima at $\x^\alpha_0=(\rr_0,\tt_0)$ of the energy boundary function $E_{\rm b}(\rr,\tt)$ \citep{Arnold:1978:book:}. It is useful to rewrite the Hamiltonian in the form
\beq
 H = H_D + H_P = \frac{1}{2} g^{rr} \p_r^2 + \frac{1}{2} g^{\theta\theta} \p_\theta^2 + H_P(r,\theta)
\eeq 
where we split $H$ into the "dynamical" $H_{\rm D}$ and the "potential" $H_{\rm P}$ parts. Introducing a small parameter $\epsilon << 1$, we can rescale coordinates and momenta by the relations 
\beq
 \x^\alpha = \x^\alpha_0 + \epsilon \hx^\alpha, \quad \p_\alpha = \epsilon \hp_\alpha,
\eeq
applied for the coordinates $\alpha \in \{ \rr,\tt \}$. We can make polynomial expansion of the Hamiltonian into the Taylor series and express it in separated parts according to the power of $\epsilon$
\bea
 H (\hp_\alpha,\hx^\alpha) &=& H_0 + \epsilon H_1(\hx^\alpha) + \epsilon^2 H_2(\hp_\alpha,\hx^\alpha) \nonumber\\
  && + \epsilon^3 H_3(\hp_\alpha,\hx^\alpha) + \ldots, \label{expand1} 
\eea
where $H_k$ is a homogeneous part of the Hamiltonian of degree $k$ considered for the momenta $\hp_\alpha$ and coordinates $\hx^\alpha$. Recall that $\p_\alpha$ occurs in the quadratic form in  (\ref{HamHam}) and appears in $H_k$ only for $k \geq 2$. If the string loop is located at a local minimum of the $E_{\rm b}(x,y)$ function, we have $H_{\rm D} = 0$ and hence $H_0 = 0$. The local extrema of the $E_{\rm b}$ function, given by (\ref{extr_a1}-\ref{extr_a2}), imply also $H_1(\hx^\alpha) = 0$.

We can divide (\ref{expand1}) by the factor $\epsilon^2$ (remember $H=0$) expressing the Hamiltonian in the vicinity of the local minimum in the "regular" plus "perturbation" form 
\beq
 H =  H_2(\hp_\alpha,\hx^\alpha) + \epsilon H_3(\hp_\alpha,\hx^\alpha) + \ldots
\eeq
If $\epsilon = 0$, we arrive to an integrable Hamiltonian
\beq
 H = H_2(\hp_\alpha,\hx^\alpha) = \frac{1}{2} \sum_\alpha \left[ g^{\alpha\alpha}(\hp_\alpha)^2 + \tilde{\omega}^2_\alpha (\hx^\alpha)^2 \right]
\eeq
representing two uncoupled harmonic oscillators. This "perturbation" approach corresponds to the linearization of the motion equations (\ref{Heq1}-\ref{Heq2}) in the neighborhood of local minima of the function $E_{\rm b}(r,\theta)$.

For the string loop motion represented by coordinates $\rr = \rr_0 + \dr,\tt = \tt_0 + \dt$ we obtain the periodic harmonic oscillations determined by the equations 
\beq
 \ddot{\dr} + \omega^2_{\rr} \, \dr = 0, \quad \ddot{\dt} + \omega^2_{\tt} \, \dt = 0,
\eeq
where the locally measured frequencies of the oscillatory motion are given by 
\beq
 \omega^2_\rr  =  \frac{1}{g_{rr}} \, \frac{\partial^2 H_P}{\partial r^2}, \quad 
 \omega^2_\tt  =  \frac{1}{g_{\tt\tt}} \, \frac{\partial^2 H_P}{\partial \theta^2}.
\eeq

According to the Kolmogorov-Arnold-Moser (KAM) theory \citep{Arnold:1978:book:}, a string loop  will oscillate in a quasi-periodic motion, if the parameter $\epsilon$ remains small. As the parameter $\epsilon$ grows, the condition $\epsilon << 1$ becomes violated, the nonlinear parts in the Hamiltonian become stronger, and the string loop enters the nonlinear, chaotic regime of its motion. Increase of non-linearity and chaoticity of a system moving in vicinity of its local stable equilibrium point is caused by increase of its energy. The transition from the regular to the chaotic regime of the motion is the solution to the "focusing" problem of the string loop trajectories discussed in \cite{Jac-Sot:2009:PHYSR4:}.

We demonstrate successive transfer from the purely regular, periodic motion through quasi-periodic motion to purely chaotic motion of a string loop in Figs. \ref{FIG_chaP1},\ref{FIG_chaP2}. The Poincare surface sections in the phase space and the Fourier transforms of the oscillatory motion in the radial and latitudinal direction clearly represent the transfer to the chaotic motion. Of course, in the entering phase of the motion with lowest energy, the string loop motion is fully regular and periodic and is represented by appropriate Lissajousse figures. 

It is convenient to represent the transfer to the chaotic system by an appropriate Lyapunov coefficient. The chaotic systems are sensitive to initial conditions and we can follow two string loop trajectories separated at the initial time $t_0$ by a small phase-space distance $d_0$. As the system evolves, the two orbits will be separated at an exponential rate if the motion of the string loops is in the chaotic regime. The Lyapunov exponent \citep{Ott:1993:book:}
\beq
\lambda_{\rm L} = \lim_{d_0 \rightarrow 0 \atop t  \rightarrow \infty } \left( \frac{1}{t} \ln \left( \frac{d(t)}{d_0} \right) \right)
\eeq
is describing the two orbits separation and hence the measure of chaos. The transition from the regular to the chaotic regime of the string loop motion is clearly visible due to the evolution of the maximal Lyapunov exponent \citep{Ott:1993:book:} demonstrated in Fig.\ref{FIG_Lyap}. We clearly see strongly increasing measure of chaos with increasing energy of the moving string loop when some critical energy is crossed. We think that such an effect is genuine to the dynamical systems and we observed it also for the string loops in the spherically symmetric braneworld spacetimes \cite{Stu-Kol:2012:JCAP:}. 

The transition from the regular to the chaotic regime of the motion is the solution to the "focusing" problem of the string trajectories discussed in \cite{Jac-Sot:2009:PHYSR4:}. The string trajectories for lower energies appear to be "focused", but this is only the effect of regularity of the system. For larger energies, string loop will be in chaotic regime, and we do not observe any string "focusing", see Figs. \ref{FIG_chaP1}, \ref{FIG_chaP2} and description there.


\begin{figure*}[htp]
\subfigure[\quad \Schw{} BH]{\includegraphics[width=0.29\hsize]{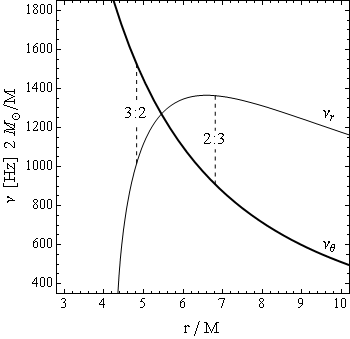}}
\hspace{0.01\hsize}
\subfigure[\quad Kerr BH $a=0.99$]{\includegraphics[width=0.29\hsize]{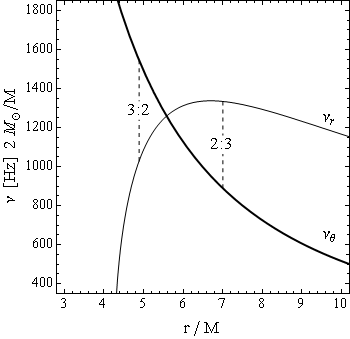}}
\hspace{0.01\hsize}
\subfigure[\quad Kerr NS $a=2$]{\includegraphics[width=0.29\hsize]{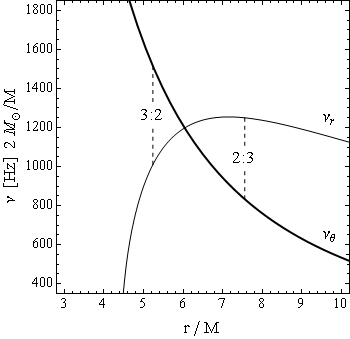}}
\caption{Radial profiles of fundamental radial and vertical frequencies $\nu_r = \Omega_r / (2 \pi), \nu_\theta = \Omega_\theta / (2 \pi)$ measured by distant observers, determined for harmonic string loop oscillations around stable equatorial equilibrium states. The string loops are considered with the parameter $\omega=0$, oscillating around compact object with mass $2M_{\rm sun}$. The loci where resonances with the ratios  $3:2$ and $2:3$ of the frequencies $\nu_\theta:\nu_r$ occur are denoted. These could be  relevant for explanation of the HF QPOs observed in microquasars \citep{Tor-etal:2011:ASTRA:, Tor-etal:2005:ASTRA:}. Both fundamental frequencies $\nu_r, \nu_\theta$ and loci of their their resonance ratios are only sightly depending on the Kerr spacetime spin parameter $a$.
\label{FIGstringQPOs}
}
\end{figure*}

\section{Quasi-periodic oscillation of string loops}

The quasi-periodic character of the motion of string loops trapped in a toroidal space around the equatorial plane of a black hole (naked singularity) suggests another interesting astrophysical application, related to the high-frequency quasi-periodic oscillations (HF QPOs) of X-ray brightness that had been observed in many Galactic Low Mass X-Ray Binaries containing neutron~stars \citep[see, e.g.,][]{Bar-Oli-Mil:2005:MONNR:,Bel-Men-Hom:2005:ASTRA:,Bel-Men-Hom:2007:MONNR:BriNSQPOCor} or black holes \citep[see, e.g.,][]{Rem:2005:ASTRN:,Rem-McCli:2006:ARASTRA:,McCli-etal:2011:CLAQG:}. Some of the~HF~QPOs come in pairs of the~upper and lower frequencies ($\nu_{\mathrm{U}}$, $\nu_{\mathrm{L}}$) of {\it twin peaks} in the~Fourier power spectra. Since the~peaks of high frequencies are close to the~orbital frequency of the~marginally stable circular orbit representing the~inner edge of Keplerian discs orbiting black holes (or neutron~stars), the~strong gravity effects have to be relevant in explaining HF~QPOs. Usually, the Keplerian orbital and epicyclic (radial and latitudinal) frequencies of geodetical circular motion are assumed in models explaining the HF QPOs in both black hole and neutron star systems \cite{Tor-etal:2005:ASTRA:,Stu-Kot-Tor:2011:ASTRA:}. However, neither of the models is able to explain the HF QPOs in all the microquasars \cite{Tor-etal:2011:ASTRA:}. Therefore, it is of some relevance to let the string loop oscillations, characterized by their radial and vertical (latitudinal) frequencies, to enter the play, as these frequencies are comparable to the epicyclic geodetical frequencies, but slightly different, enabling thus some corrections to the predictions of the models based on the geodetical epicyclic frequencies. Of course the frequencies of the string loop oscillations in physical units have to be related to distant observers. For the string loops with $\omega=0$, the radial and latitudinal oscillatory frequencies take the form
\begin{widetext}
\bea
 \Omega^2_{\rm r} &=& \left( \frac{c^3}{GM} \right)^2 \frac{3 a^6 r+a^4 (r (2 r (5 r-9)+9)-6)-a^2 r^3 (r ((r-10) r+35)-27)-r^6 ((r-5) r+3)}{r \left(a^2-r^3\right) \left(a^2 (r+2)+r^3\right)^2}, \label{QPOs1} \\
\Omega^2_{\rm \theta} &=& \left( \frac{c^3}{GM} \right)^2 \frac{a^4 (2-3 r)+2 a^2 (3-2 r) r^2-r^5}{r^2 \left(a^2-r^3\right) \left(a^2 (r+2)+r^3\right)} \label{QPOs2}
\eea
\end{widetext}
We demonstrate their dependence on the radial coordinate of the equilibrium position of the string loop and on the black hole (naked singularity) spin in Fig. \ref{FIGstringQPOs}. It is quite interesting that the latitudinal oscillatory frequency of the string loop for \Schw{} BH ($a=0$) equals to the latitudinal frequency of the epicyclic geodetical motion as observed by distant observers - for details see \cite{Stu-Kol:2012:JCAP:}. The radial and latitudinal frequencies of the string loop oscillations and the geodetical epicyclic motion in Kerr spacetimes are different, enabling thus substantial changes of the relation of the frequency ratio for HF QPOs modeled by the string loop motion and the geodetical epicyclic oscillations.

\section{Conclusions}

Scalar field $\varphi$ living on the string loops is crucial for creating the centrifugal forces, governed by the angular momentum parameter $J$, and hence for the existence of stable equilibrium positions. String loop equilibrium positions are located  between photon $r_{\rm{ph}}$ and marginally stable $r_{\rm ms}$ test particle orbits, supporting thus the view of string loop model as a composition of charged particles and the related electromagnetic field \cite{Cre-Stu:2013:PhRvE:}. 

Contrary to the motion of current-carrying string loops in the spherically symmetric spacetimes that is of a degenerate type, depending on the magnitude of the angular momentum of the string loop $J$ only \cite{Kol-Stu:2010:PHYSR4:,Stu-Kol:2012:PHYSR4:,Stu-Kol:2012:JCAP:}, in the Kerr spacetimes the situation is more complex, since the string loop motion depends also on the "axial" component of the angular momentum described by the parameter $\omega$. This situation is similar to those related to the test particle motion, see, e.g., \cite{Mis-Tho-Whe:1973:Gra:}, and enlarges substantially the possible signatures of astrophysical effects occurring in the strong gravity in Kerr spacetimes. In the present paper we have focused our study on the phenomena demonstrating clear distinction between the BH and NS spacetimes and on the role of the string loop parameter $\omega$ in astrophysically relevant phenomena. 

Various types of the string loop dynamics expressed by the constant energy sections of the energy boundary function $E=E_{\rm b}(x,y)$ can be observed in the Kerr BH and Kerr NS spacetimes. Only four different types of the energy boundary sections can exist in the BH spacetimes \cite{Jac-Sot:2009:PHYSR4:}; the absence of the horizon in the NS spacetimes leads to new types of energy boundary function sections whose behavior strongly changes the character of the string loop motion: off-equatorial minima and additional equatorial minima located close to the ring singularity arise in this case. Note that string loop oscillations occuring near the off-equatorial minima can serve as an important signature of the Kerr NS spacetimes, enabling clear distinguishing of the Kerr BH spacetimes. 

Large string acceleration along the $y$-axis can occurs only for large $E/2J$ ratios. The transmutational accelerating effect is only slightly depending on the Kerr black hole spin $a$, but there is clear distinction between BH and NS cases; Kerr NS are definitely better accelerators and the spin dependence is in this case stronger than in the black hole case - region with strong gravity around origin of coordinates is now accessible, see Fig. \ref{stringFIG_9}. Also the role of the string loop parameter $\omega$ is clearly demonstrated in our results. 

Trajectory "focusing" problem, introduced in \cite{Jac-Sot:2009:PHYSR4:}, can be clearly explained as an effect of transition from regular to the chaotic regime of motion due to the energy increase. Such transitions are observed in many other dynamical system \cite{Kop-Kar-Kov-Stu:2010:ASTRJ:,Sem-Suk:2012:MNRAS:}, and are not a special property of the string loop model.

Small string-loop quasi-periodic oscillations around stable equilibrium positions, governed by Eqs (\ref{QPOs1}-\ref{QPOs2}), are potentially relevant to HF QPOs observed in binary systems containing  black holes or neutron stars and will be studied in a forthcoming work. Our preliminary results indicate that the HF QPOs observed in the four microquasars discussed in \cite{Tor-etal:2011:ASTRA:}, could be explained by the model of the string loop oscillations due to the role of the additional string loop parameter $\omega$.

\section{Acknowledgments}

The authors would like to express their acknowledgments for the Institutional support of the Faculty of Philosophy and Science of the Silesian University at Opava, the internal student grant of the Silesian University SGS/23/2013, EU grant Synergy CZ.1.07/2.3.00/20.0071 and GA{\v C}R 202/09/0772.


 
\end{document}